\documentclass[11pt,a4paper]{article}
\pdfoutput=1 

\usepackage{jcappub}

\usepackage[T1]{fontenc} 
\usepackage{amsmath,amssymb,gensymb}
\usepackage{xspace}
\usepackage{siunitx,booktabs}

\newcommand{\wimpsim}{\textsf{WimpSim}\xspace}
\newcommand{\wimpann}{\textsf{WimpAnn}\xspace}
\newcommand{\wimpevent}{\textsf{WimpEvent}\xspace}
\newcommand{\pythia}{\textsf{Pythia-6.4.26}\xspace}
\newcommand{\darksusy}{\textsf{DarkSUSY-6}\xspace}

\DeclareSIUnit{\au}{AU}
\DeclareSIUnit{\gev}{GeV}
\DeclareSIUnit{\tev}{TeV}

\usepackage[dvipsnames]{xcolor}

\newcommand{\ovr}{\overline}
\newcommand{\fermi}{\emph{Fermi}-LAT}

\title{Neutrinos and gamma rays from long-lived mediator decays in the Sun}

\author[a]{Carl Niblaeus,}
\author[a,b,1]{Ankit Beniwal\note{ORCID ID: \href{https://orcid.org/0000-0003-4849-0611}{0000-0003-4849-0611}}}
\author[a]{and Joakim Edsj\"{o}}

\affiliation[a]{The Oskar Klein Centre for Cosmoparticle Physics, Department of Physics, \\
Stockholm University, AlbaNova, SE-106 91 Stockholm, Sweden}
\affiliation[b]{Center for Cosmology, Particle Physics and Phenomenology (CP3), \\
Universit\'{e} catholique de Louvain, B-1348 Louvain-la-Neuve, Belgium}

\emailAdd{carl.niblaeus@fysik.su.se}
\emailAdd{ankit.beniwal@uclouvain.be}
\emailAdd{edsjo@fysik.su.se}

\abstract{We investigate a scenario where dark matter (DM) particles can be captured and accumulate in the Sun, and subsequently annihilate into a pair of long-lived mediators. These mediators can decay further out in the Sun or outside of the Sun. Compared to the standard scenario where DM particles annihilate directly into Standard Model particles close to the solar core, here we also obtain fluxes of gamma rays and charged cosmic rays. We simulate this scenario using a full three-dimensional model of the Sun, and include interactions and neutrino oscillations. In particular, we perform a model-independent study of the complementarity between neutrino and gamma ray fluxes by comparing the recent searches from IceCube, Super-Kamiokande, \textit{Fermi}-LAT, ARGO and HAWC.

We find that the resulting neutrino fluxes are significantly higher at high energy when the mediators decay further out in the Sun. We also find that gamma ray searches place stronger constraints than neutrino searches on these models even in cases where the mediators decay mainly inside the Sun, except in the approximately inner 10\% of the Sun where neutrino searches are more powerful. We present our results in a model-independent manner and release a new version of the \textsf{WimpSim} code that can be used to simulate this scenario for arbitrary mediator models.}


\arxivnumber{1903.11363}

\subheader{CP3-19-09}

\begin{document} 

\maketitle

\flushbottom

\section{Introduction}
Dark Matter (DM) accounts for about 85\% of the total matter density in our Universe \cite{Aghanim:2018eyx}.~Yet its particle nature and properties remain elusive despite a wealth of astrophysical/cosmological evidence to support its existence. Although a recent discovery of the long-awaited Higgs boson \cite{Aad:2012tfa, Chatrchyan:2012xdj} has completed the Standard Model (SM) of particle physics, it has been unsuccessful in offering viable particle DM candidates.~Thus, theorists have focused instead on studying theories beyond the SM, e.g., supersymmetry \cite{Martin:1997ns}. One popular class of particle DM candidates include Weakly Interacting Massive Particles (WIMPs) \cite{Bertone:2004pz}. Not only do they appear naturally in well-motivated theories beyond the SM, e.g., supersymmetric models with a neutralino WIMP, a GeV--TeV scale WIMP with a weak-scale interaction cross-section can accurately reproduce the observed DM abundance. For recent reviews, see refs.~\cite{Tanabashi:2018oca,Roszkowski:2017nbc,Arcadi:2017kky}.

To probe the particle nature and properties of DM as well as their potential interactions with ordinary matter, various detection strategies are employed.~These range from \emph{direct searches} \cite{Penning:2017tmb, Liu:2017drf, Undagoitia:2015gya} for an elastic scattering of DM particles off nucleons in deep underground laboratories (e.g., PandaX-II, LUX, XENON1T), \emph{indirect searches} \cite{Conrad:2017pms,Slatyer:2017sev,Gaskins:2016cha, Danninger:2017wjw} for signs of DM annihilation in high-density regions of the Universe (e.g., Galactic Centre, dwarf spheroidal galaxies, Sun) using ground/space-based telescopes (e.g., H.E.S.S., \fermi, HAWC, AMS-02, IceCube), and direct production of DM particles at the LHC (\emph{collider searches}) \cite{Boveia:2018yeb,Felcini:2018osp,Kahlhoefer:2017dnp,Buchmueller:2017qhf}.~All of these provide complementary limits on the allowed particle DM models. 

In our local DM halo, DM particles can be gravitationally captured by the Sun and accumulate at its centre \cite{Press:1985ug,Silk:1985ax}.~The capture of DM particles is initiated by a scattering process where they lose enough energy to fall below the escape velocity of the Sun, and thus become gravitationally bound to it. In subsequent scatterings with the solar material, they lose more energy and eventually sink to the centre of the gravitational well where they can thermalise and annihilate into SM particles.~As both the scattering leading to capture and the annihilation depend on cross sections of a DM model, searching for annihilation products from the Sun can probe the particle nature of DM.

In general, DM particles can annihilate into various SM particles that can decay or hadronise into stable decay products, e.g., gamma rays, neutrinos and/or charged particles. Out of these, only neutrinos can escape the dense core of the Sun and give rise to an observable DM signal that can be detected using Earth-based telescopes.~In this respect, neutrino telescopes offer a unique way of searching for a DM signal from the Sun. However, neutrinos from DM annihilation near the solar core have to pass through much of the solar material on their way out of the Sun.~Due to the neutral-current (NC) and charged-current (CC) neutrino-nucleon interactions, the neutrino fluxes are significantly attenuated for energies above $\sim 100$\,GeV. Thus, traditional solar DM searches at neutrino telescopes \cite{Aartsen:2016zhm} are limited to neutrino energies in the range 1--1000 GeV.

\begin{figure}[t]
    \centering
    \includegraphics[scale=1.4]{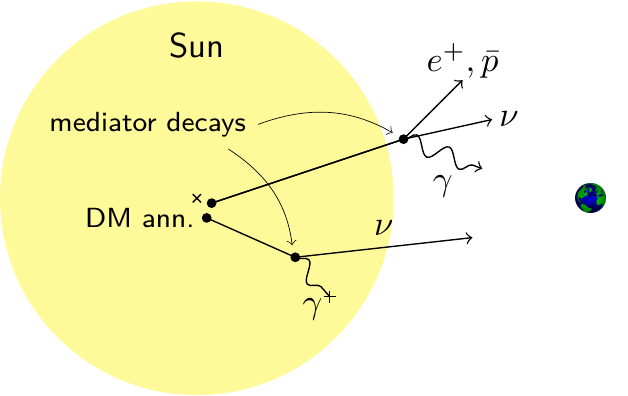}
    \caption{An illustration of how the solar DM searches are affected when DM particles do not annihilate directly into SM particles but rather into a pair of long-lived mediators which decay further away from the solar core. Decays away from the centre reduce the impact of neutrino interactions with the solar materials. In particular, decays outside the Sun opens up a new window to searches for gamma rays and charged cosmic rays. }
    \label{fig:long_lived_case}
\end{figure}

In this paper, we will investigate a scenario that goes beyond the usual WIMP scenario by assuming that our DM particles $\chi$ do not annihilate directly into SM particles, but instead into long-lived mediators $Y$ that subsequently decay into SM particles, see figure~\ref{fig:long_lived_case} for an illustration.~We typically get this phenomenology in secluded DM models \cite{Pospelov:2007mp,Pospelov:2008jd,Batell:2009zp,Fortes:2015qka,Okawa:2016wrr,Yamamoto:2017ypv,Siqueira:2019wdg}.~This scenario has been studied in more general scenarios for the Sun in refs.~\cite{Bell:2011sn,Arina:2017sng,Leane:2017vag,Albert:2018jwh,Adrian-Martinez:2016ujo,Ardid:2017lry}, but here we will extend previous analyses and provide a tool for future studies.

With long lifetimes, the mediators can propagate away from the point of DM annihilation in the Sun before they decay.~As the solar DM density falls (approximately) exponentially with the solar radius, the resulting neutrino fluxes suffer from much less attenuation due to CC interactions than in the standard WIMP scenario.~This leads to an enhancement of the neutrino flux \cite{Bell:2011sn} that extends up to higher neutrino energies even for decay lengths shorter than the solar radius.~This enhancement has major implications for neutrino telescopes as their sensitivity approximately scales as the square of the neutrino energy.~In addition, for mediator decay lengths comparable to or larger than the solar radius, gamma rays and charged cosmic rays can also escape the Sun and be observed at the Earth.~In the standard scenario, these final states are not observable as they are immediately absorbed by the solar material after their production in the Sun.~Thus, a measurement of a positive signal in the direction of the Sun by both neutrino and gamma ray/cosmic-ray detectors can point towards the long-lived mediator scenario. 

A specific realisation of the long-lived mediator scenario in the context of DM models is the dark photon model \cite{Agrawal:2018vin,Kouvaris:2016ltf,Cirelli:2016rnw,Fortes:2015qka}.~In these models, DM particles can be collected at the centre of the Earth/Sun or stars \cite{Brdar:2016ifs} where they can annihilate into dark photon pairs. These dark photons can reach the surface of the Earth/Sun where their decays can lead to observable fluxes of gamma rays, neutrinos and charged cosmic rays.~Recent studies have shown that some parts of the model parameter space can be explored using the current generation of neutrino and space-based cosmic-ray instruments \cite{Feng:2015hja,Feng:2016ijc,Ardid:2017lry}.~In fact, the proposed sensitivity of these searches can probe new parts of the parameter space compared to beam dump experiments \cite{Kahn:2014sra,Andreas:2012mt,Gninenko:2012eq}, Big Bang Nucleosynthesis \cite{Berger:2016vxi}, 1987A supernova cooling \cite{Hardy:2016kme,Chang:2016ntp,Rrapaj:2015wgs} and direct DM searches.

In this paper, we will work in a model-independent setup instead of focusing on concrete long-lived mediator models.~We will assume that our DM particle physics model contains a DM particle $\chi$ that annihilate into a long-lived mediator $Y$.~We assume that there are processes (either through $Y$ exchange or other processes) that facilitate scatterings of $\chi$ in the Sun so that we get capture and annihilation of $\chi$ in the Sun.~We will then set up a simulation framework where we allow for the (boosted) mediator $Y$ to decay further out in the Sun or outside of the Sun, and produce neutrinos, gamma rays and charged cosmic rays. For neutrinos, we include interactions (both NC and CC scatterings) and oscillations in a full three-flavor setup. Our simulation tools are released as a new version of the \wimpsim code \cite{Blennow:2007tw, Edsjo:2017kjk, wimpsimonline}, version 5.0, that can be used for phenomenological work or as an event generator for experimental studies.

We will here focus on the relation between the gamma ray and neutrino fluxes, and leave the charged particle fluxes for a future study.~The charged particles are affected by magnetic fields and hence propagation uncertainties, warranting a propagation model that falls outside the scope of the current article.~Using the mediator mass, DM mass, boosted mediator decay length and decay channel as the relevant parameters of interest, we show the effects of their variation on the resulting muon neutrino and gamma ray fluxes at the Earth. In situations where a sizeable gamma ray signal is expected, we also test the complementarity between the neutrino and gamma ray signals.~This allows us to identify regions in the parameter space where these searches can provide the strongest sensitivity to the long-lived mediator scenario.

Limits on the long-lived mediator scenario from solar DM searches using neutrinos have previously been placed by the ANTARES collaboration \cite{Adrian-Martinez:2016ujo} and from the IceCube 79-string data \cite{Ardid:2017lry}.~In both studies they focus on mediator decays that mainly lead to a neutrino signal.~As we are primarily interested in the complementarity with a gamma ray signal, we focus mainly on decay channels that give rise to significant amounts of neutrinos and gamma rays. We will however also investigate the muon channel studied in refs.~\cite{Adrian-Martinez:2016ujo,Ardid:2017lry} as even in this scenario, we expect some gamma rays due to final state radiation from the $Y \rightarrow \mu^+ \mu^-$ decay. Other limits come from ref.~\cite{Leane:2017vag} where both gamma ray and neutrino signals are considered, and ref.~\cite{Albert:2018jwh} where HAWC limits on the solar gamma ray flux are used. In comparison to these studies, we complement the analysis, in particular by considering other values for the mediator decay length (where our inclusion of the effect of neutrino interactions in the Sun is crucial for short decay lengths), and by investigating the complementarity between gamma ray and neutrino searches more explicitly. 

The rest of the paper is organised as follows.~In section~\ref{sec:model}, we define the relevant model parameters, and outline key observations and constraints for the long-lived mediator scenario.~After giving a brief review of the standard \wimpsim code in section~\ref{sec:method}, we discuss our new additions for the long-lived mediator scenario. Sections~\ref{sec:simulation} and \ref{sec:analysis} describe how we perform our simulations and extract neutrino and gamma ray limits on the DM annihilation rate respectively.~Our final results such as the effects of varying the boosted mediator decay length and its mass on the muon neutrino fluxes, and complementarity between gamma ray and neutrino searches are presented in section~\ref{sec:results}.~We conclude in section~\ref{sec:conclusions}. In Appendix~\ref{app:nulimits}, we provide more details on our method for extracting the neutrino telescope limits in the long-lived mediator scenario.

\section{Model definition and constraints}\label{sec:model}

\subsection {Model definition}
Using a phenomenological approach, we do not model the details of the dark sector. Instead, we focus on the parameters that are needed to obtain the spectra of neutrinos and gamma rays from the long-lived mediator decays, and to compare them against experimental searches. In a study of a specific particle physics scenario, one can constrain the fundamental parameters by mapping them onto the ones we use in this study.

\begin{table}[t]
    \centering
    \begin{tabular}{cc}
        \toprule \toprule
         Parameters & Description \\ \midrule
         $m_\chi$ & Dark Matter (DM) mass \\[1.5mm]
         $m_Y$ & Mediator mass \\[1.5mm]
         $\gamma L$ & Boosted mediator decay length \\[1.5mm]
         -- & Mediator decay channel \\[1.5mm]
         $\Gamma_A$ & DM annihilation rate in the Sun \\ \bottomrule \bottomrule
    \end{tabular}
    \caption{Parameters used in our analysis to describe the spectra of neutrinos and gamma rays from long-lived mediator decays.}
    \label{tab:modelparams}
\end{table}

In table~\ref{tab:modelparams}, we show the parameters that we use in our study. The DM mass $m_\chi$ sets the absolute scale of the spectrum and limits the maximal energy in a DM annihilation. The main importance of the mediator mass $m_Y$ is to determine the kinematically allowed decay channels --- it has a smaller impact on the spectrum shape than the other parameters, as we will show in section~\ref{sec:results}.~We parametrise the mediator lifetime by the parameter $\gamma L$, i.e., the Lorentz boosted vacuum decay length $L$, defined as
\begin{align}
    \gamma L = \gamma c\tau_0 = (m_\chi/m_Y) c\tau_0,
\end{align}
where $\tau_0$ is the vacuum lifetime and $\gamma=m_\chi/m_Y$ is the Lorentz boost of the mediator for DM annihilations at rest.~For the mass combinations that we are interested in, the mediators are (in general) highly boosted when they emerge from the DM annihilation.~Thus, the combination $\gamma L$ is a phenomenologically more relevant parameter to use than the vacuum lifetime or decay length. We assume that the mediators decay with a 100\% branching ratio into a chosen SM decay channel of interest. As the decay channel changes the spectrum of neutrinos and gamma rays rather significantly, we also leave it as a free parameter in our analysis.

The above parameters are sufficient to obtain the spectra of neutrinos and other particles of interest per annihilation. To compare against experimental searches, we must also know the DM annihilation rate in the Sun, denoted by $\Gamma_A$. This rate can in principle be calculated from DM scattering and annihilation cross sections by analysing how DM particles are captured in the Sun.~We do not consider the scattering and annihilation cross sections in this study. Instead, we assume that for a given set of model parameters, there exists some combination of cross sections that results in a DM annihilation signal from the Sun with rate $\Gamma_A$.

\subsection{Relevant observations and constraints}
We will consider below the limits that the experimental data sets on the annihilation rate of DM particles in the Sun in the long-lived mediator scenario.~As we focus on neutrinos and gamma rays, we use data from the IceCube telescope \cite{Aartsen:2016zhm} and Super-Kamiokande \cite{Choi:2015ara} for the case of neutrinos, and observations from \fermi~\cite{Abdo:2011xn,Tang:2018wqp}, HAWC \cite{Albert:2018jwh,Albert:2018vcq} and ARGO \cite{Bartoli:2019xvu} for gamma rays. 

The existence of a high energy solar gamma ray flux is now well established from analysis of EGRET and \emph{Fermi}-LAT data  \cite{Orlando:2008uk, Abdo:2011xn,Linden:2018exo, Tang:2018wqp}.~Apart from a possible DM signal, the observed flux at high energy is expected to be composed of two known mechanisms that are backgrounds for a DM search: one where gamma rays are produced in the cascades formed when cosmic rays hit the outer parts of the Sun \cite{Seckel:1991ffa,Zhou:2016ljf}, and another where gamma rays are produced in inverse Compton scattering by cosmic ray electrons on sunlight \cite{Moskalenko:2006ta,Orlando:2006zs, Orlando:2017iyc}. The former mechanism also produces a neutrino flux \cite{Seckel:1991ffa,Edsjo:2017kjk,Arguelles:2017eao,Ng:2017aur} which will act as a background for neutrinos from DM annihilations.~If a high energy neutrino signal from the Sun is also observed, the relationship between the neutrino and gamma ray fluxes can potentially be used to distinguish a DM signal from these backgrounds. 

The observed flux of gamma rays from the Sun is however not completely understood: the flux is significantly larger than the available theoretical predictions and in addition, it possesses unexpected spectral properties \cite{Tang:2018wqp,Nisa:2019mpb}. It is expected that the solar magnetic field impacts the gamma ray signal from cosmic ray interactions by mirroring some of the cosmic ray cascades  back towards the Earth, which may explain the discrepancy.~Nevertheless, the explanation of the observed flux is currently incomplete. 

The mediator lifetime is also constrained by Big Bang Nucleosynthesis (BBN) as a large lifetime can lead to prohibitively high levels of energy injection in the thermal plasma \cite{Jedamzik:2006xz}. In our results, we show a region excluded for a BBN limit of $\tau^* = \SI{1}{\second}$. We caution however that the BBN limit on the lifetime varies greatly depending on the decay channel and abundance of the mediators in the early Universe. An assessment of the abundance, in particular, requires a knowledge of the DM and mediator production mechanism which is beyond the scope of the general approach we take in this study. The BBN limit constrains the parameters in our setup as
\begin{equation}
    \gamma L = \frac{m_\chi}{m_Y} c\tau_0 \lesssim \frac{m_\chi}{m_Y} c \tau^*,
\end{equation}
where $\tau^*$ is an upper limit on the mediator lifetime from BBN. This expression can be rewritten in terms of the DM mass, the mediator mass and its boosted decay length as
\begin{equation}\label{eq:lim_BBN}
    m_\chi \gtrsim 2.32  \, m_Y \, \frac{\gamma L}{R_\odot} \, \left( \frac{\SI{1}{\second}}{\tau^*}\right).
\end{equation}
In section~\ref{sec:results}, we will indicate this limit in our plots assuming $\tau^*=\SI{1}{\second}$. 

In principle, the long-lived mediator scenario can be constrained by searches for long-lived particles at the LHC. However, as we are primarily interested in scenarios where the decay length is large, we expect the coupling to SM particles to be small and hence the production cross sections at the LHC to also be small and escape detection. 

Other limits on the long-lived mediator scenario can come from indirect DM searches, such as searching for effects of DM annihilation on the cosmic microwave background radiation or for gamma rays from astrophysical sources \cite{Profumo:2017obk}.~In the scenario we consider in this paper, we work in a model-independent framework where neither the annihilation nor the scattering cross section is specified. Although we will set limits on the DM annihilation rate in the Sun, $\Gamma_A$, we cannot compare to or set limits on the annihilation cross section since the relationship between $\Gamma_A$ and the annihilation cross section is not one-to-one, but depends on the solar DM capture rate, which we do not consider here.

\section{Methodology} \label{sec:method}
The \wimpsim\ code can simulate DM annihilations in the Sun, take care of neutrino interactions and oscillations on their way out of the Sun and to the Earth, and simulate neutrino interactions at a neutrino telescope. Here we will use this code, but update it to version 5.0 to also handle our long-lived mediator scenario.~We will include neutrino interactions and oscillations, a fully three-dimensional model of the Sun and \pythia\ \cite{Sjostrand:2006za} for particle decays and hadronisation.~We briefly review what the \wimpsim\ code can do and then describe our new simulation framework.

\subsection{Review of the \wimpsim code}\label{subsec:wimpsim_refresh}
Here we describe the standard \wimpsim\ code \cite{Blennow:2007tw,Edsjo:2017kjk,wimpsimonline} for general DM annihilations.~In the next subsection, we address the modifications done to handle the case of long-lived mediators. 

\wimpsim\ is a Fortran code that simulates neutrinos from DM annihilations in the Sun and the Earth, where \pythia\ is used to simulate the DM annihilation and the decay and/or hadronisation of the annihilation products.~The annihilation point is taken from a probability distribution which assumes that the DM follows a thermal distribution (which has recently been shown in ref.~\cite{Widmark:2017yvd} to be a very good approximation).~For the Sun, it then takes care of propagation of neutrinos out of the Sun, and includes CC and NC interactions, and oscillations (in a full three-flavour setup) \cite{Blennow:2007tw}. In case a CC interaction takes place that produces a tau lepton, it then lets the tau lepton decay and includes the resulting neutrinos in the subsequent propagation. It can then propagate the neutrinos to the Earth (including vacuum oscillations), let them interact near the detector and produce muons (or other leptons and hadronic showers) that can be detected by a neutrino telescope such as IceCube or Super-Kamiokande.~A similar procedure is handled for annihilations in the Earth.

\wimpsim\ is fully event based and follows each neutrino from the creation to the interaction near the detector.~It generates output in the form of summary tables (suitable for further phenomenological studies) and event files (suitable for neutrino telescope Monte Carlos). For the summary files, the end result is very similar to the results in ref.~\cite{Cirelli:2005gh}.~To be flexible, \wimpsim is split up into two parts, \wimpann\ that takes care of the DM annihilations and propagation out of the Sun and to 1 AU, and \wimpevent\ that takes care of the final propagation and interactions near the detector. In this latter program, one can create simulations for detectors at different locations on the Earth and for different time periods.

In an earlier study \cite{Edsjo:2017kjk}, some of us (CN,\;JE) investigated neutrinos from cosmic ray interactions in the Sun which resulted in a modification to \wimpsim\ by an addition of a new module to be run before \wimpevent. In the current paper, we adopt a similar approach by adding a new module for long-lived mediator decays. 

\subsection{Simulation of the long-lived mediator scenario}
\label{sec:simmed}
To make it possible to simulate long-lived mediator decays in \wimpsim, we have included a new particle that represents the mediator with a mass and decay channel as specified by the user. For the long-lived mediator scenario, we then call a different event generation routine in \wimpsim that generates two boosted mediator decays for each DM annihilation.

\begin{figure}
    \centering
    \includegraphics[scale=1.4]{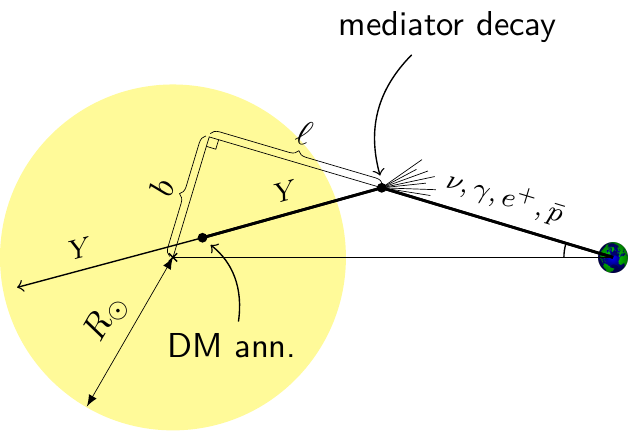}
    \caption{Geometry of the long-lived mediator decays. DM particles annihilate into two long-lived mediators (denoted $Y$) near the solar core. The mediators are emitted back-to-back and propagate out from the core before decaying (inside or outside the solar surface) into pairs of SM particles. The resulting decay chain contains neutrinos and other particles of interest that travel towards the Earth.~We obtain fluxes of these particles using our code. The impact parameter $b$ indicates how far out the particle from the mediator decay is from the Earth-Sun axis (in units of $R_{\odot}$). For instance, $b=1$ gives a path that just grazes the solar edge, whereas $b=0$ is a path through the solar core. The length $\ell$ specifies how far along the path (defined by $b$) the particle is as measured from the line that goes straight out from the core such that $\ell=0$ is the point in the upper left (right angle) corner of the triangle in this figure. }
    \label{fig:geom}
\end{figure}

In figure~\ref{fig:geom}, we show the geometry of our setup. The DM particles annihilate at a point near the solar centre into two long-lived mediators which travel some distance before they decay into a pair of SM particles.~We parametrise the path of these particles from the mediator decays with coordinates $b$ and $\ell$ where $b$ is the impact parameter (i.e., the perpendicular distance from the solar centre to the path), and $\ell$ is the length along that path as measured from a point straight outwards from the centre (the upper left corner of the triangle in figure~\ref{fig:geom}). Thus, $\ell$ can be positive or negative, but it always increases as particles propagate towards the Earth. 

Relative to their production at the DM annihilation point, we let the mediators propagate a distance before they decay, with this distance sampled from an exponential decay distribution
\begin{equation} \label{eqn:decay_prob}
    P(\ell_Y) = \frac{1}{\gamma L} e^{-\ell_Y/\gamma L},
\end{equation}
where $\gamma L$ is the boosted decay length of the mediator and $\ell_Y$ is the length travelled by the mediator from the annihilation point.~We assume that the mediators propagate without scattering on the solar material.~At the decay point (inside or outside the Sun), we let the boosted mediator decay into a decay channel specified by the user and use \pythia to simulate the decay. In each mediator decay, we collect all neutrinos and store them as individual events. We do not assume that neutrinos travel along the path of the mediator (although this is a good approximation for large boost factors) and take into account the spreading around the mediator direction.~This is however only relevant in isolated cases, e.g., when $m_\chi\sim m_Y$ such that $\gamma\sim1$, so that the decay products are emitted with large separation angles. 

The spreading around the mediator direction will have the effect that neutrinos encounter a different integrated column density when exiting the Sun compared to the one-dimensional approximation. This changes the amount of absorption and hence the observed neutrino flux.

Apart from changing the impact of interactions, the spreading will also diffuse the neutrino and gamma ray signals, and increase the apparent size of the source.\footnote{Although it is not implemented in the current release of \wimpsim, a functionality to print out two-dimensional histograms in energy and impact parameter (or angle from the Earth-Sun axis) is straightforward to add if there is an interest in the community for such a feature.}~This effect becomes larger for smaller boost of the mediator -- large boosts lead to a more collimated, point-like signal. A proper inclusion of this reduces the experimental sensitivity as the signal is spread out and more background is included. Neutrino telescopes are likely more affected than gamma ray telescopes, given the lower angular resolution. For most parts of the parameter regions that we are concerned with here, the signal is confined to small angles and comes mainly from within the solar disk.

In some cases, mediators will decay at a position further away from the Sun than \SI{1}{\au}, resulting in particles that cannot be seen at Earth.~In particular, the fraction of such decays for large $\gamma L$ is significant. In our simulations, we do not keep particles that come from mediator decays happening further away than \SI{1}{\au}. 

Hadrons containing heavy quarks travel some distance and lose energy before they decay. Thus, if a neutrino is formed in the decay of a hadron containing bottom or charm quarks, we reduce its energy according to the treatment outlined in ref.~\cite{Ritz:1987mh}. As the interaction lengths depend on the density of the solar material at the point where these hadrons are produced, we scale the interaction lengths properly to account for mediator decays happening away from the solar centre. We also make sure that pions, kaons and muons formed in mediator decays happening outside the Sun decay (inside the Sun, these are quickly stopped and only give rise to low-energy neutrinos which we are not interested in). The resulting neutrinos are then allowed to interact and oscillate on their way to Earth as per the standard treatment in \wimpsim (see subsection~\ref{subsec:wimpsim_refresh}).

For mediator decays outside the Sun, a visible signal of gamma rays and charged cosmic rays can be produced. We construct histograms in kinetic energy for positrons, antiprotons and gamma rays coming from decays outside the solar surface (i.e., we do not save these particles as individual events). Antiprotons often come from decays of antineutrons which for most energies have a boosted decay length larger than \SI{1}{\au}. Thus, they travel a significant distance from the point of the mediator decay before their decay. Due to this fact, we require that antiprotons are produced before \SI{1}{\au}. For charged particles, magnetic effects from the Interplanetary Magnetic Field are not taken into account. In this study, we focus primarily on neutrinos and gamma rays, and leave the study of charged cosmic rays from long-lived mediator decays for future work. 

As a validation of our implementation, we have compared against the standard DM annihilation scenario, for which \wimpsim has been extensively tested. In the standard scenario, DM particles annihilate into short-lived particles. In the long-lived mediator scenario, we can recover the standard scenario by taking the limit of $\gamma L\to 0$ and $\gamma \to 1$ (i.e., $m_\chi=m_Y$), ending up with mediators decaying at rest at the DM annihilation point. However, as the kinematics are different in $\chi \chi \to 2\,\mathrm{SM}$ and $\chi\chi\to Y Y \to (2\,\mathrm{SM})\,(2\,\mathrm{SM})$, one also has to change the DM mass by a factor of 2. Furthermore, as there are two mediator decays per DM annihilation, one also has to divide the resulting flux by a factor of 2 when taking this limit to obtain the same neutrino flux.~We have checked that the simulated neutrino fluxes are indeed identical in this limit.

\section{Simulation setup} \label{sec:simulation}
In this section, we summarise the various settings used in our numerical simulations of the long-lived mediator scenario.~Apart from numerical values of the parameters in table~\ref{tab:modelparams}, these include our choice for the neutrino oscillation parameters, solar model and the neutrino detector properties. 

In our simulations, we use for the neutrino oscillation parameters the best-fit values  from \textsf{NuFit-3.2} \cite{Esteban:2016qun, nufitonline}; these are also summarised in table~\ref{tab:osc-params}.~We only consider normal mass hierarchy; a detailed study of different oscillation scenarios is beyond the scope of our study. 

\begin{table}[t]
    \centering
    \begin{tabular}{cc}
        \toprule \toprule
         Parameters & Best-fit value  \\ \midrule
         $\theta_{12}$ & 33.62\degree \\[1.5mm]
         $\theta_{13}$ & 8.54\degree \\[1.5mm]
         $\theta_{23}$ & 47.2\degree \\[1.5mm]
         $\delta_{\rm{CP}}$ & 234\degree \\[1.5mm]
         $\Delta m_{21}^2$ & \SI{7.40e-5}{\electronvolt\squared}   \\[1.5mm]
         $\Delta m_{31}^2 $ & \SI{2.494e-3}{\electronvolt\squared} \\ \bottomrule \bottomrule
    \end{tabular}
    \caption{Neutrino oscillation parameters used in our simulations. These are based on the best-fit values from \textsf{NuFit-3.2} \cite{Esteban:2016qun} in the normal mass hierarchy.}
    \label{tab:osc-params}
\end{table}

The mass differences give rise to oscillations as the neutrinos propagate from their production at the Sun to the Earth.~We can define oscillation lengths $\lambda_{ij}$ that determine the characteristic length of these oscillations, one for each of the $\Delta m_{ij}^2$.~In the approximation that only vacuum oscillations are relevant, these lengths are given by 
\begin{equation}\label{eqn:osc-lengths}
    \frac{\lambda_{21}}{\SI{1}{\au}} \approx 0.22 \left( \frac{E}{\SI{1}{\tev}} \right) , \quad \frac{\lambda_{3i}}{\SI{1}{\au}} = \num{6.6e-3} \left( \frac{E}{\SI{1}{\tev}} \right),\quad i=1,\;2,
\end{equation}
where effectively we have two oscillations lengths (as $\Delta m_{32}^2\approx \Delta m_{31}^2$).~Broadly speaking, we expect neutrino oscillations to be visible in the spectra when $\lambda_{ij}\lesssim \SI{1}{\au}$, which defines a characteristic energy scale for each $\lambda_{ij}$ that marks this transition. There is also a transition at lower energies, below which oscillations are washed out in the spectra when $\lambda_{ij}$ becomes on the order of the size of the production region. For each $\lambda_{ij}$, we can then separate into three energy ranges: energies low enough that $\lambda_{ij}$ is comparable to the size of the neutrino production region and oscillations wash out; intermediate energies where oscillations are visible in the spectra; and high energies so that the $\lambda_{ij}$ are large compared to \SI{1}{\au}, and oscillations do not develop during the propagation between the Sun and the Earth. 

\begin{table}[t]
    \centering
    \begin{tabular}{llccc}
        \toprule \toprule
        Scenario & Decay channel & $m_\chi$\,(GeV) & $m_Y$\,(GeV) &  $\gamma \emph{L}/\emph{R}_\odot$ \\ \midrule
        A) Varying $\gamma L/R_\odot$ & $Y \to b \overline{b}$ & $1000$ & $20$ &[0.01,\;10] \\[1.5mm] 
        & $Y \to \tau^+ \tau^-$ & $1000$ & $20$
        &[0.01,\;10] \\ \midrule
        B) Varying $m_Y$ & $Y \to b \overline{b}$ & 5000 & $\{20,\;200,\;2000\}$ & 1 \\[1.5mm]
        & $Y \to \tau^+ \tau^-$ & 5000 & $\{20,\;200,\;2000\}$ & 1 \\ \midrule
        C) 3D treatment & $Y \to \tau^+ \tau^-$ & $5000$ & $4900$ & $0.3$ \\ \midrule
        D) $\gamma$-$\nu$ comparison & $Y \to \tau^+ \tau^-$ & $[100,\; 10^4]$ & $20$ & $[0.01,\;10]$ \\[1.5mm]
        & $Y \to b \ovr{b}$ & $[100,\; 10^4]$ &  $20$ & $[0.01,\;10]$ \\[1.5mm]
        & $Y \to \tau^+ \tau^-$ & $[10,\; 10^4]$ & $4$ & $[0.01,\;10]$ \\ \midrule
        E) Muon channel & $Y \to \mu^+ \mu^-$ & $[10,\; 10^4]$ & $1$ & $[0.01,\;10^6]$ \\
        \bottomrule \bottomrule
    \end{tabular}
    \caption{Simulation parameters for studying the effect of varying $\gamma L/R_\odot$ where $\gamma \equiv m_\chi/m_Y = 50$ (Scenario A), the effect of varying the mediator mass $m_Y$ (Scenario B), the effect of using a full three-dimensional (3D) treatment of the mediator decay geometry (Scenario C), the gamma ray-neutrino comparison (Scenario D), and the muon channel to compare against a previous work \cite{Ardid:2017lry} (Scenario E).}
    \label{tab:sim_setup}
\end{table}

For the neutrino interactions and oscillations we need a solar model. In our analysis, we use the default solar model in \darksusy, which is the one by Serenelli et al. \cite{Serenelli:2009yc}.

In table~\ref{tab:sim_setup}, we summarise our choice for the simulation parameters used to study various scenarios. Each of these scenarios are described in more detail below.
\begin{description}
    \item[Varying $\gamma L/R_\odot$ (Scenario A):] In this scenario, we study the effect of different mediator boosted decay lengths $\gamma L$ on the muon neutrino and gamma ray fluxes at 1 AU. This is done for both a soft ($b \ovr{b}$) and hard ($\tau^+ \tau^-$) decay channel. A DM mass of 1000\,GeV is chosen to illustrate the impact of CC and NC interactions with the solar material on the neutrino fluxes. Thus, for large $\gamma L$ values, a higher neutrino flux is expected as neutrinos suffer less absorption and interactions on their way to the solar surface.
    
    \item[Varying $m_Y$ (Scenario B):] This scenario is used to test the importance of the mediator mass $m_Y$ on the muon neutrino and gamma ray fluxes at 1 AU. The simulation parameter used are based on ref.~\cite{Leane:2017vag}. There it was pointed out that the mediator mass has a small effect on the gamma-ray fluxes. We check this point for both the muon neutrino and gamma ray fluxes.
    
    \item[3D treatment (Scenario C):] In all our simulations, we take into account the fact that the decay products of the mediator do not follow exactly the trajectory of the decaying mediator. To test the effect of the 3D treatment relative to the 1D approximation where one assumes the same trajectory for the mediator and the decay products, we simulate the same scenario with our standard 3D treatment and the 1D approximation. As the relative angle between the mediator and its decay products is on average larger for a smaller boost $\gamma$ of the mediator, we study a scenario where $\gamma = m_\chi/m_Y \sim 1$. The spread in the neutrino direction in the 3D treatment causes the neutrinos to pass through a different amount of solar matter as compared to the one-dimensional approximation. As we expect a large difference in the case of higher neutrino absorption, we choose a relatively large $m_\chi$, let the mediator decay to $\tau^+\tau^-$ and choose a relatively small value for $\gamma L$.\footnote{Although it would lead to more absorption, a too small value for $\gamma L$ results in a smaller difference between the three- and one-dimensional calculations, as the decay products would in both cases, regardless of the angle relative to the mediator, follow approximately radial trajectories out of the Sun and thus encounter similar amounts of material in the two cases.}
    
    \item[$\gamma$-$\nu$ comparison (Scenario D):] As we make comparisons between gamma ray and neutrino searches, we also made a set of special simulations for this purpose. For this scenario, we focus on the range of parameters where we expect to see interesting effects, with two mediator masses ($m_Y = 20$\,GeV, $4$\,GeV) and for each one a wide range of DM masses and $\gamma L$ values. We have chosen points more or less evenly distributed in the logarithm of $m_\chi$ and $\gamma L$ but with a slightly denser grid around $\gamma L \simeq 0.1\,R_\odot$. 

    \item[Muon channel (Scenario E):] This scenario is very similar to scenario D, except that we focus on lighter mediator masses and mediator decays to $\mu^+ \mu^-$. This simulation goes up to higher $\gamma L$ values as we want to compare our results against those in ref.~\cite{Ardid:2017lry}. 
\end{description}

For scenarios A--C, we generate neutrino and gamma ray fluxes at 1 AU from the Sun, i.e.,\ the average Sun-Earth distance. For scenarios D--E, when we compare against neutrino searches, we propagate the neutrinos all the way to the detector and generate neutrino-nucleon interactions at the detector. For simplicity, these simulations are performed for a detector at the South pole, with ice as the detector medium and averaged over the austral summer. For Super-Kamiokande, this introduces a small error of the order of 10\%, which is smaller than errors from other approximations used in our study. In all of our simulations, we have included final state radiation from charged particles, which will prove to be important especially in scenario E.

We also made special scans for regular WIMP particles that cover the same set of WIMP model parameters as in the IceCube \cite{Aartsen:2016zhm} and Super-Kamiokande \cite{Choi:2015ara} analyses. These are used to be able to relate their limits on neutrinos from WIMPs to more general DM (see Appendix~\ref{app:nulimits}). For simplicity, we also ran these simulations for a detector at the South pole, averaging over the austral summer.

\section{Analysis of neutrino and gamma ray limits}\label{sec:analysis}
Here we describe our method used to extract upper limits on the DM annihilation rate $\Gamma_A$ using the data from neutrino and gamma ray telescopes.~This allows us to show the complementarity between neutrino and gamma ray searches in parts of the parameter space. 
 
To study the complementarity between neutrino and gamma ray searches, we look at a region in the ($\gamma L,\;m_\chi$) plane for fixed mediator masses $m_Y$. For each ($\gamma L,\;m_\chi$) point in this plane, we calculate the current limit on the DM annihilation rate from neutrino and gamma ray searches respectively.~For neutrinos, we use the results from IceCube \cite{Aartsen:2016zhm} and Super-Kamiokande \cite{Choi:2015ara}, whereas for gamma rays, we use observations from \fermi\; \cite{Abdo:2011xn,Tang:2018wqp}, HAWC \cite{Albert:2018jwh,Albert:2018vcq} and ARGO \cite{Bartoli:2019xvu}. 

The spectrum of a neutrino signal from long-lived mediator decays can differ quite significantly from a spectrum in the standard DM annihilation scenario.~Thus, we use the method outlined in Appendix~\ref{app:nulimits} to generalise the published limits on DM annihilation in the Sun to also be applicable for other spectra, like our mediator decay spectra considered here. For IceCube, we expect this method to be good to about a factor of 2, whereas for Super-Kamiokande we expect the method to be good to about 25\%.~Even though a proper full analysis by the experimental collaborations should be able to produce more accurate limits, we will see below that this is good enough for our purpose here.~This is due to a very strong $\gamma L$ dependence (in the region of interest) of our gamma ray fluxes that makes small uncertainties in the way we extract limits less important. 

For the gamma rays, we estimate approximate limits by just assuming that the measured gamma ray flux is an upper limit on the contribution from mediator decays from the Sun. This is rather conservative as we know that there are other sources of the gamma ray flux, namely cosmic ray interactions in the solar atmosphere and inverse Compton scattering of electrons on solar photons. However, as the measured gamma ray flux from the Sun is not fully understood, we will just assume that our gamma ray flux from mediator decays cannot be larger than the measured fluxes.

For both the neutrino and gamma ray fluxes, our \wimpsim\ simulations will give us an estimate of the neutrino and gamma ray flux per annihilation.~By comparing against the limits as outlined above, we can then translate this into a limit on the DM annihilation rate in the Sun, $\Gamma_A^\nu$ and $\Gamma_A^\gamma$ respectively. For this comparison, we define the ratio
\begin{equation}\label{eqn:ratio_eta}
   \eta \equiv \Gamma_A^\nu/\Gamma_A^\gamma.
\end{equation}
The stronger the limit, the lower the respective value of $\Gamma_A$ will be. Hence, $\eta > 1$ will mean that the gamma ray limit is stronger, while $\eta < 1$ will mean that the neutrino limit is stronger. 

For the gamma ray fluxes, we can use a simple analytical scaling to convert the flux between different $\gamma L$ values. This is because the spectrum is just the fully decayed spectrum from \textsf{Pythia}, normalised to account for the probability that the decay happens between the Sun and Earth.\footnote{For the neutrino fluxes, we do not use this scaling as one gets non-zero contributions from mediator decays in the solar interior; one also needs to account for the effect of absorption.}~For a given $\gamma L$, the probability for the mediator decaying between the solar surface and \SI{1}{\au} is given by the exponential decay probability in eq.~\eqref{eqn:decay_prob} integrated between $R_\odot$ and the Earth, e.g. for $\gamma L=R_\odot$, $\sim$\,37\% of the mediators are expected to decay outside the Sun and contribute to the visible gamma ray signal.~The spectra for different $\gamma L$ values are then related by the ratio of their respective decay probabilities. Thus, one can obtain the spectrum at one value of $\gamma L$ by scaling the flux at another $\gamma L$ value with the ratio of the two probabilities. 

The above scaling neglects the small effect from the fact that DM particles annihilate in a region around the centre of the Sun, and assumes that mediators travel from the centre of the Sun and radially outwards. The size of the annihilation region depends on the DM mass but is typically confined to at most a few percent of the solar radius.~We therefore expect this to affect the scaling procedure only for $\gamma L/R_\odot$ values of the order of a few percent. However, in this case, the tail of the probability distribution that we integrate over in eq.~\eqref{eqn:decay_prob} is (in any case) so small that the gamma ray flux is negligible. 

In our figures that make use of the gamma ray limits (figures~\ref{fig:2dmed20} to~\ref{fig:icmu-2d}), we have used this method to scale our gamma ray fluxes from the $\gamma L/R_\odot = 1$ simulations to lower $\gamma L$ values. This is done in order to get better statistics at lower $\gamma L$ values. We have also verified that we get the same result (but noisier) when we compare the actual simulation data to the scaled fluxes at lower $\gamma L$ values.

When we calculate the limits, we include the full flux (i.e.,\ neglect the angular distribution of the fluxes). This is a good approximation as our signals have a small angular extent for most of the parameter values that we consider 
(see the discussion in Section~\ref{sec:simmed}).

\section{Results and discussion}\label{sec:results}
In this section, we show results from our \wimpsim simulations.~We show in the first part (sections~\ref{sec:results_gammaL},~\ref{sec:results_medmass} and~\ref{sec:results_3d}) the effects of varying the mediator boosted decay length ($\gamma L$), decay channel, mediator mass ($m_Y$) and our full three-dimensional direction of the decay products on the muon neutrino and gamma ray fluxes. We do not add together the muon neutrino and antineutrino fluxes in our plots, but show specifically the flux of muon neutrinos (the total $\nu_\mu+\overline{\nu}_\mu$ flux is approximately twice that of $\nu_\mu$ alone). As the figures in sections~\ref{sec:results_gammaL},~\ref{sec:results_medmass} and~\ref{sec:results_3d} are mainly intended to illustrate the dependence on the various parameters in our setup, we show the fluxes at the average Earth-Sun distance of 1 AU rather than the fluxes in an actual detector on Earth (where the eccentricity of the Earth's orbit and the time during the year when the observations take place would be taken into account).~We have checked that all of our parameter choices for scenarios A, B and C in table~\ref{tab:sim_setup} satisfy eq.~\eqref{eq:lim_BBN} for the BBN limit $\tau^* =\SI{1}{s}$.

We also show results in section~\ref{sec:results_nuvsga} that demonstrate the complementarity between gamma ray and neutrino signals, illustrating where in the ($\gamma L,\;m_\chi$) plane (for fixed mediator mass) a particular type of search provides the strongest sensitivity. In all cases, we consider 2 decay channels for the mediators: one with a harder energy spectrum for the neutrinos ($\tau^+ \tau^-$), and another with a softer energy spectrum ($b\ovr{b}$). As for the BBN constraint in eq.~\eqref{eq:lim_BBN}, we indicate the regions in our 2D plots where this constraint is violated. 

Finally, in section~\ref{sec:results_muon}, we show results for the decay channel $Y\to \mu^+ \mu^-$ and compare with a previous study \cite{Ardid:2017lry} of this scenario.

\subsection{Effect of varying the boosted decay length $\gamma L$}\label{sec:results_gammaL}
\begin{figure}[t]
    \centering
    \includegraphics[width=0.49\textwidth]{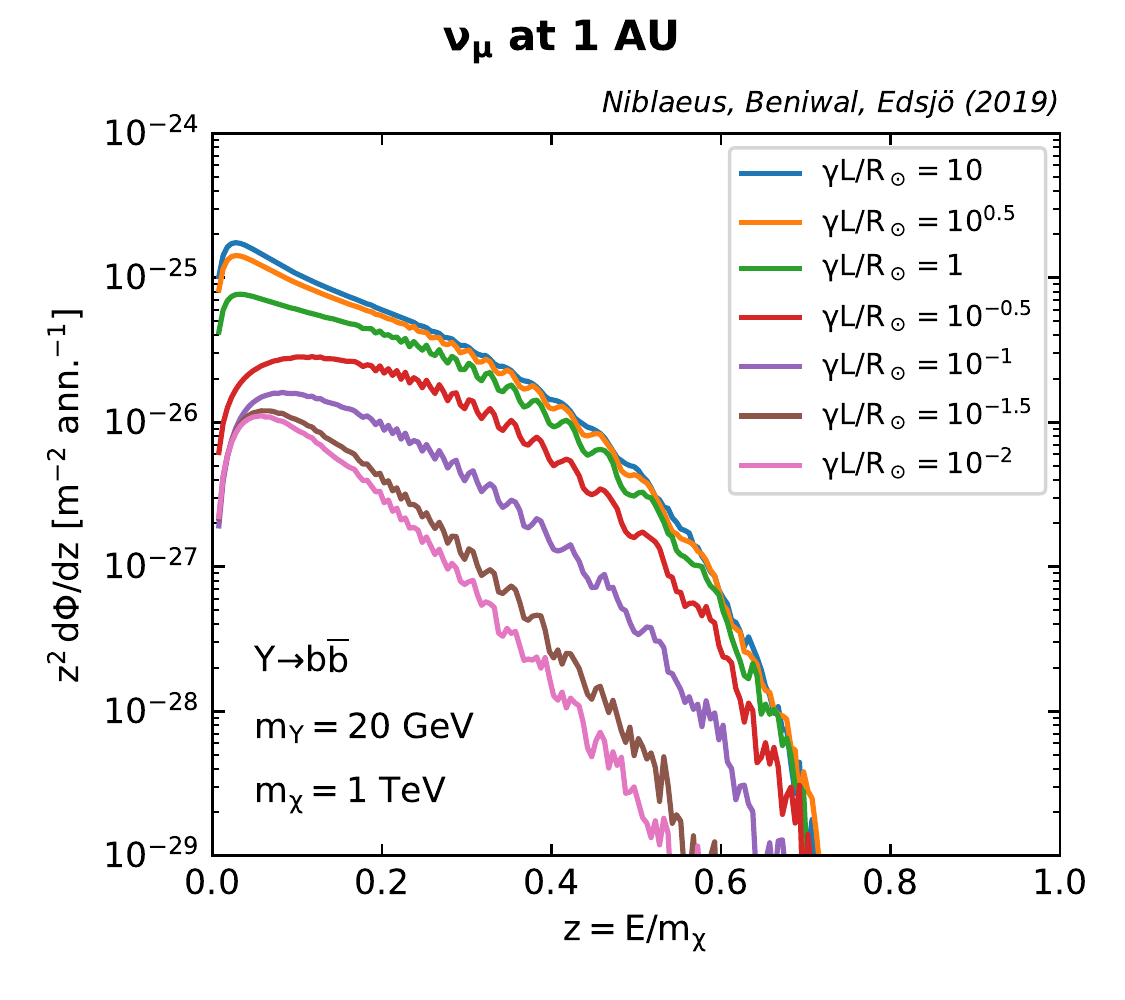}
    \includegraphics[width=0.49\textwidth]{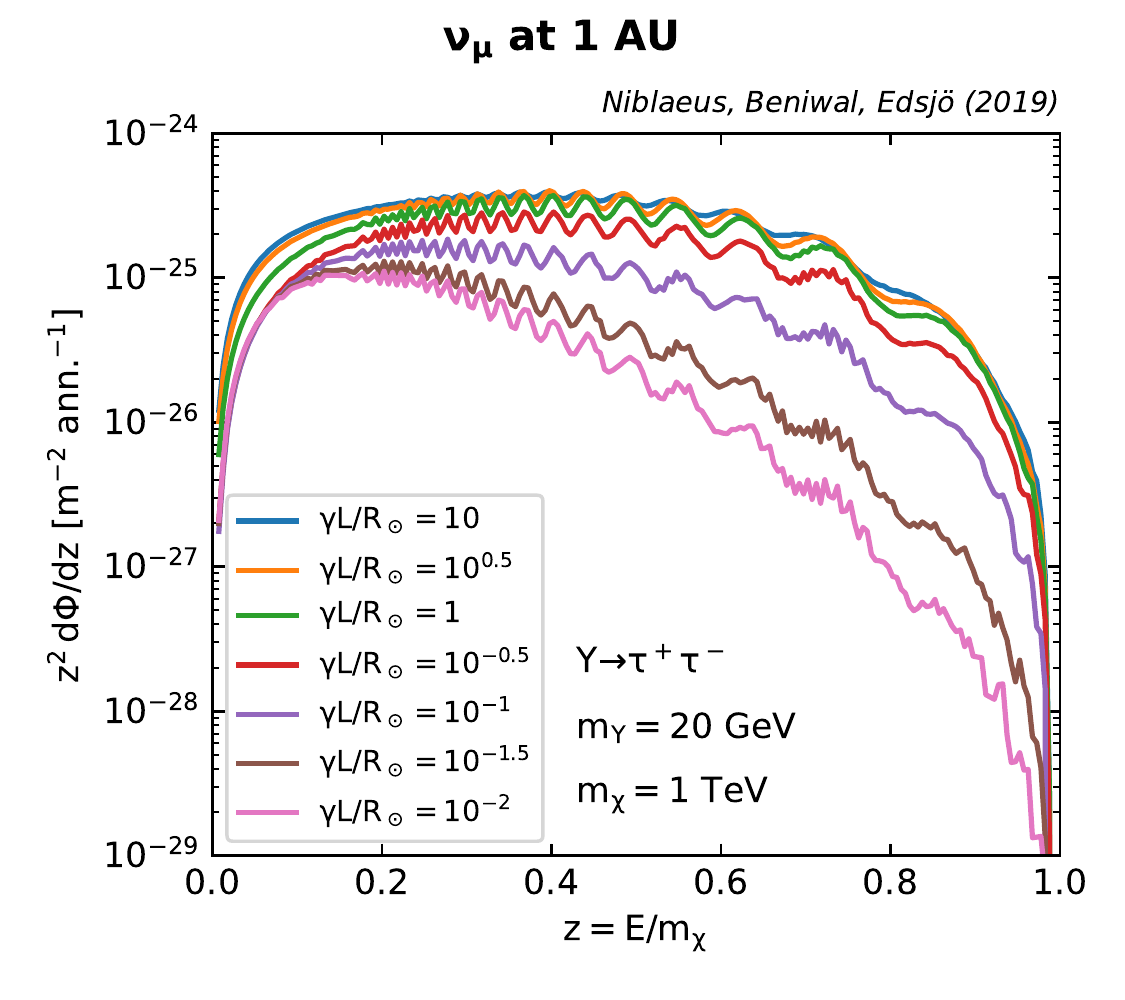}
    \vspace{-1.5em}
    \caption{Muon neutrino fluxes at 1 AU from $Y \rightarrow b\ovr{b}$ (\emph{left panel}) and $Y \rightarrow \tau^+ \tau^-$ (\emph{right panel}) for various boosted decay lengths $\gamma L$ (in units of the solar radius $R_\odot$).~Our choice for the simulation parameters are shown in table~\ref{tab:sim_setup} (Scenario A).}
    \label{fig:neutrino-changing_Ldec}
    \vspace{2.5mm}
    \includegraphics[width=0.49\textwidth]{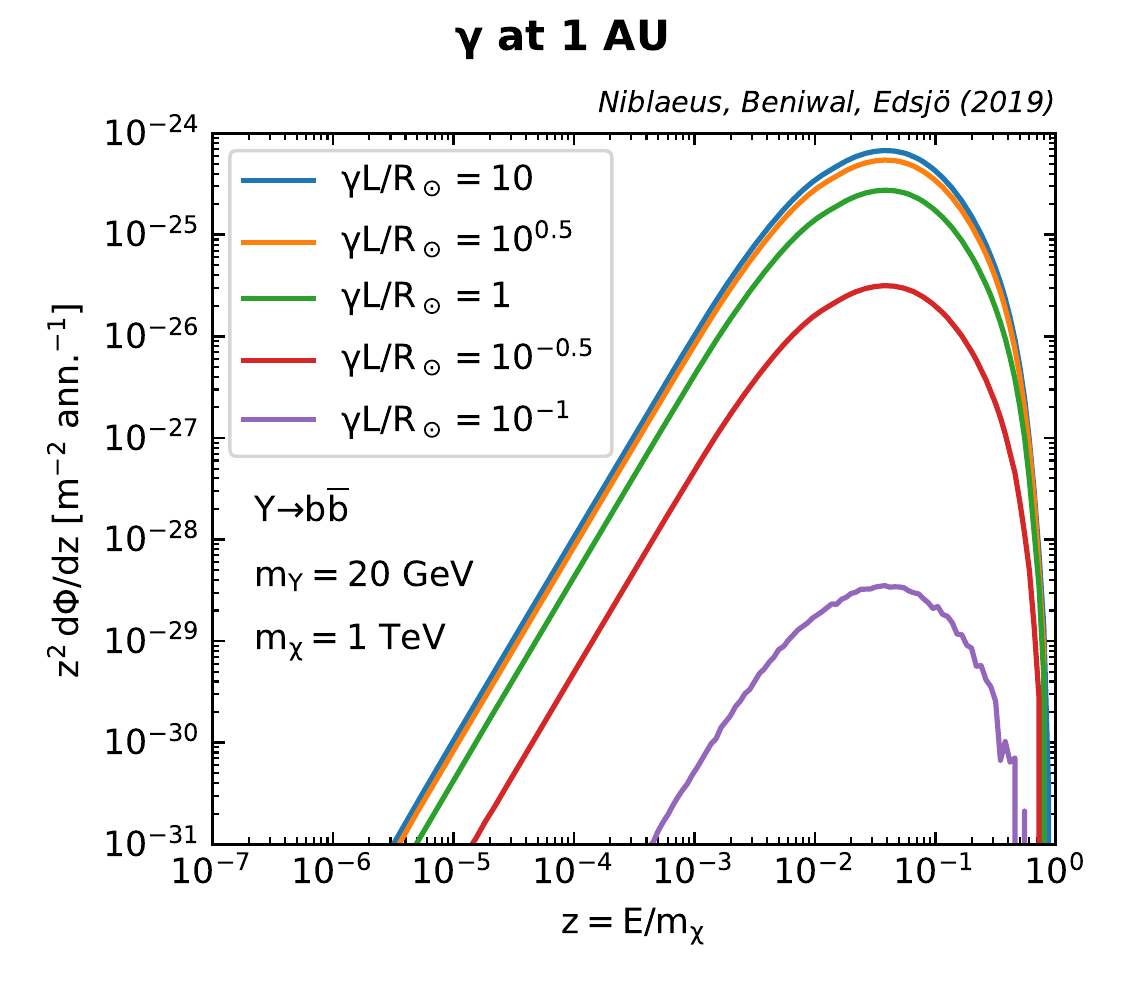}
    \includegraphics[width=0.49\textwidth]{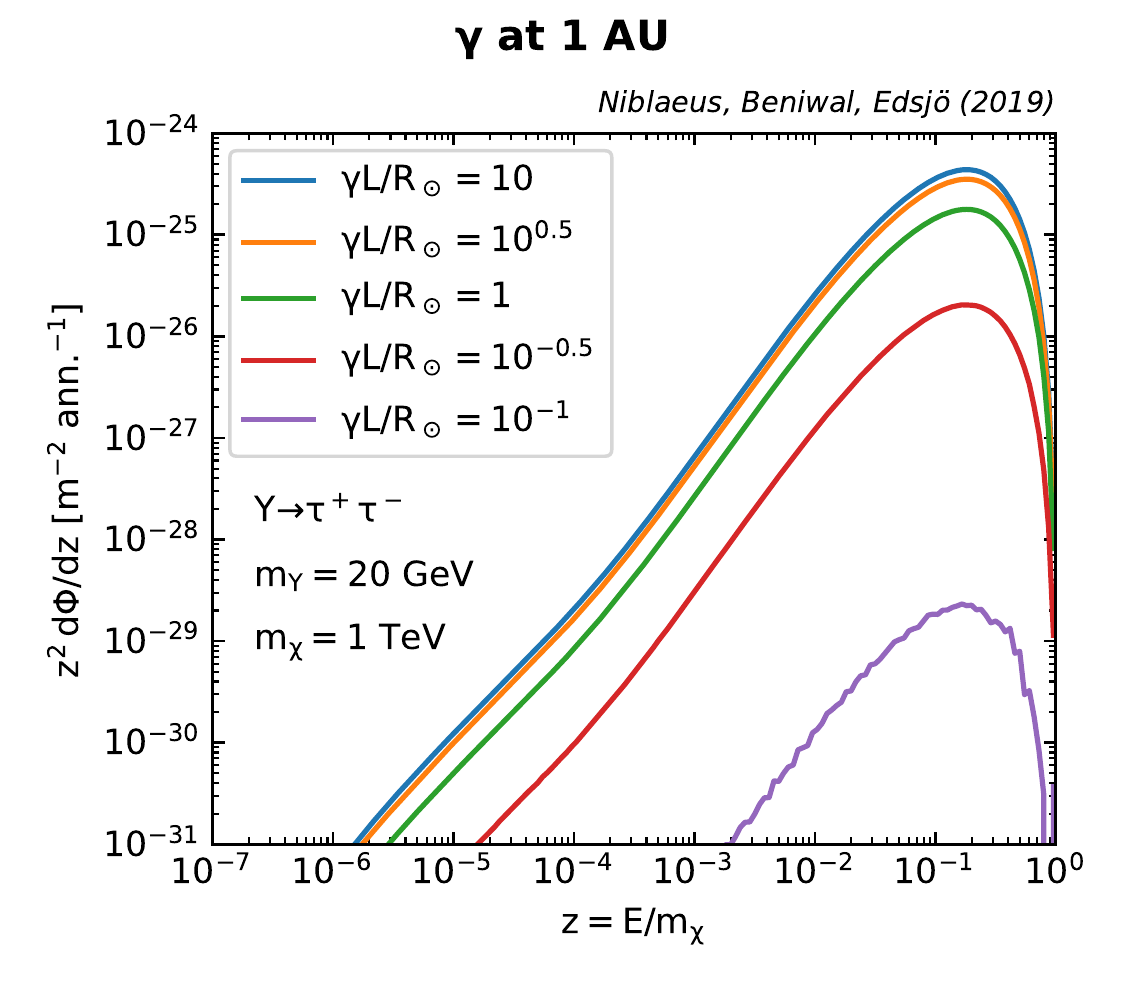}
    \vspace{-1.5em}
    \caption{Same as figure~\ref{fig:neutrino-changing_Ldec} for gamma rays fluxes.}
    \label{fig:flux-gamma-changing_Ldec}
\end{figure}

In figure~\ref{fig:neutrino-changing_Ldec}, we show the effects of varying $\gamma L$ on the muon neutrino fluxes for the parameter values corresponding to scenario A in table~\ref{tab:sim_setup}. In general, we can see that the flux increases as $\gamma L$ increases, in agreement with a previous study \cite{Bell:2011sn}.~This is expected as for larger $\gamma L$, neutrinos have a larger probability to be produced further out from the solar core and thus are subject to less absorption. 

We can also clearly see the effect of neutrino oscillations in figure~\ref{fig:neutrino-changing_Ldec}.~In terms of the oscillation lengths $\lambda_{ij}$ in eq.~\eqref{eqn:osc-lengths}, we mostly see the effect of oscillations due to $\lambda_{21}$ in this energy range, which appear clearly for $E_\nu \gtrsim$~\SI{200}{\gev}.~The oscillations due to $\lambda_{3i}$ mainly show up at higher energies and mostly average out in figure~\ref{fig:neutrino-changing_Ldec}, although they are causing the small wiggles at the highest $z$ shown.~The oscillations are dominated by the vacuum oscillations between the Sun and \SI{1}{\au}.

For comparison, we also show the corresponding gamma ray fluxes at \SI{1}{\au} in figure~\ref{fig:flux-gamma-changing_Ldec}. Again, we can see an expected rise in flux when $\gamma L$ is increased.~Although the gamma ray flux only receives a contribution from mediator decays outside the Sun, there is still a non-negligible gamma ray flux also for $\gamma L/R_\odot < 1$ as there is a non-zero probability of mediator decays happening outside the Sun.~However, for $\gamma L \lesssim 0.1\,R_\odot$, this probability is so small that the resulting gamma ray flux is negligible.

As $\gamma L$ is increased for fixed values of the remaining parameters, the flux will continue to increase until it reaches a maximum, and then for some $\gamma L/R_\odot > 1$, it will start to decrease. This is due to a competition between the increased flux from more mediators decaying outside the Sun, and a decrease in flux due to more and more decays happening further away than \SI{1}{\au}. For some value of $\gamma L$, the flux is therefore maximised.

\subsection{Effect of varying the mediator mass $m_Y$}\label{sec:results_medmass} 

\begin{figure}[t]
    \centering
    \includegraphics[width=0.49\textwidth]{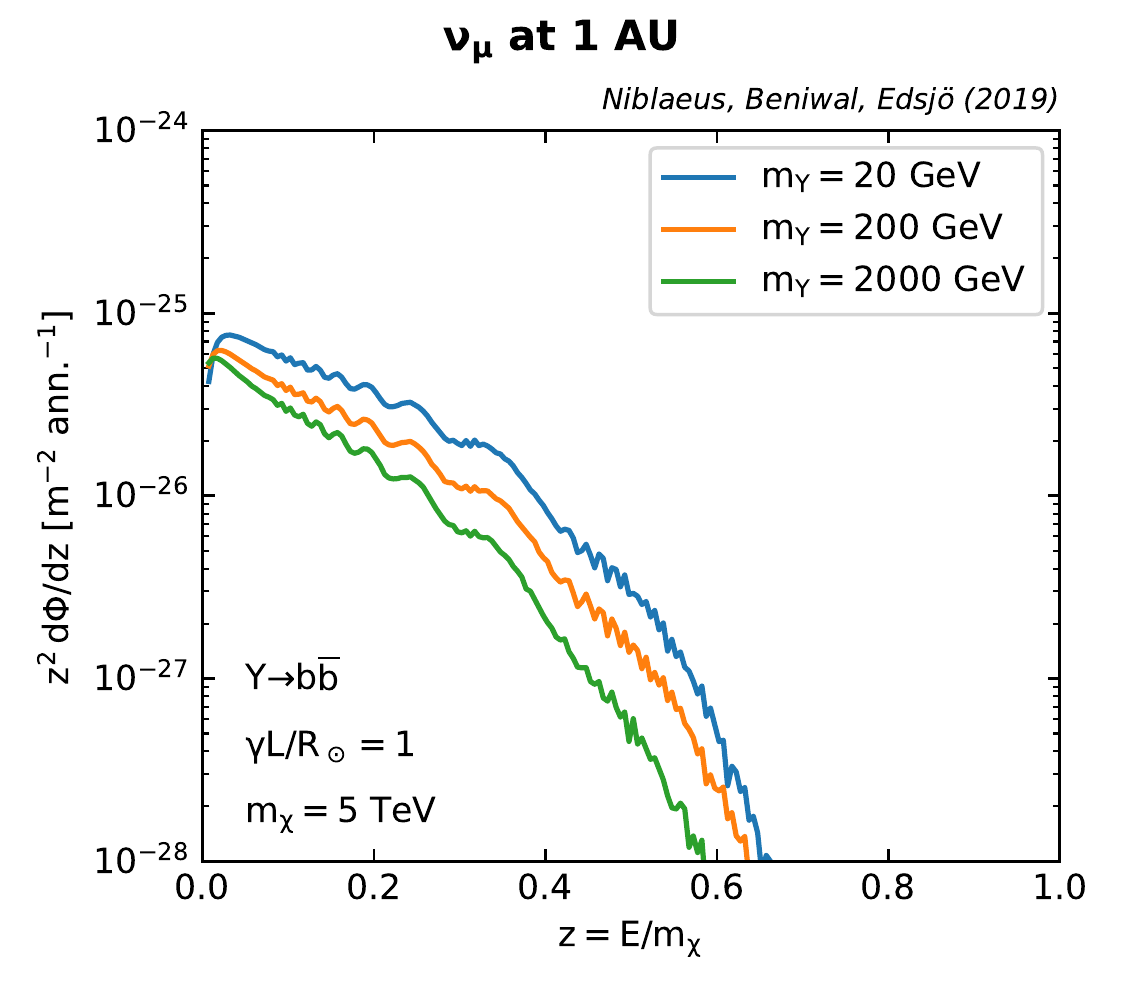}
    \includegraphics[width=0.49\textwidth]{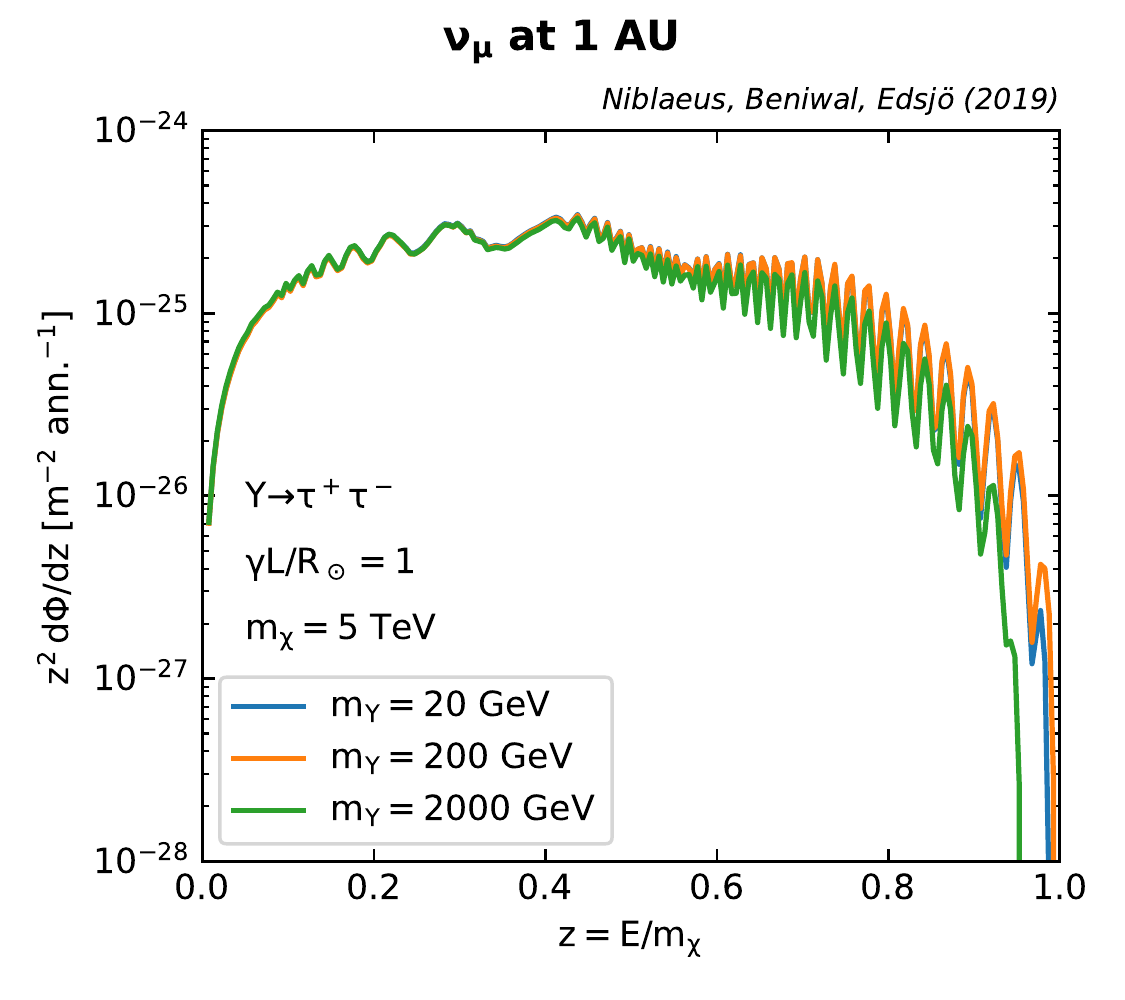}
    \caption{Muon neutrino fluxes at 1 AU from $Y \rightarrow b\ovr{b}$ (\emph{left panel}) and $Y \rightarrow \tau^+ \tau^-$ (\emph{right panel}) for $\gamma L = R_\odot$ and various mediator masses. Our choice for the mediator masses are summarised in table~\ref{tab:sim_setup} (Scenario B).}
    \label{fig:nu-flux-medmass}
    \vspace{2.5mm}
    \includegraphics[width=0.49\textwidth]{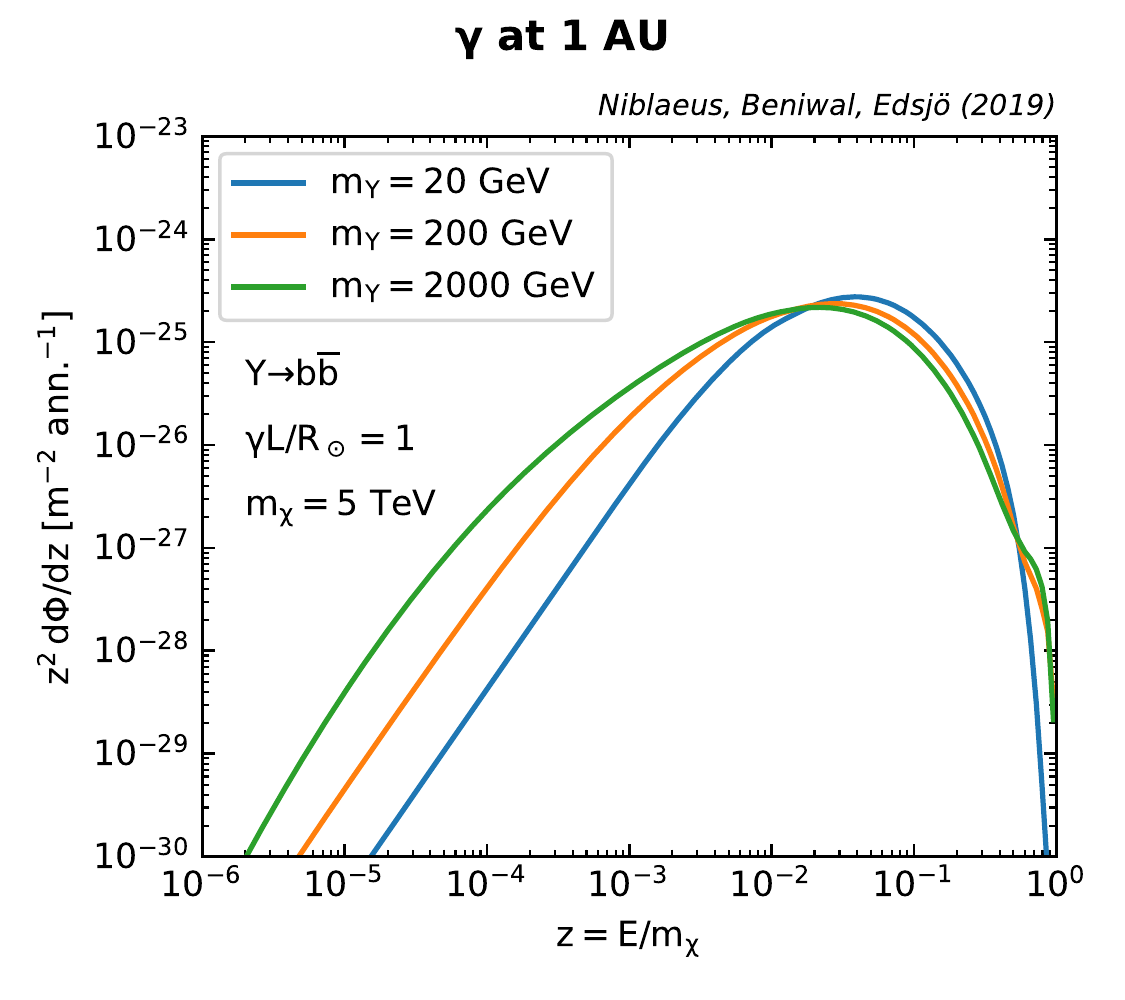}
    \includegraphics[width=0.49\textwidth]{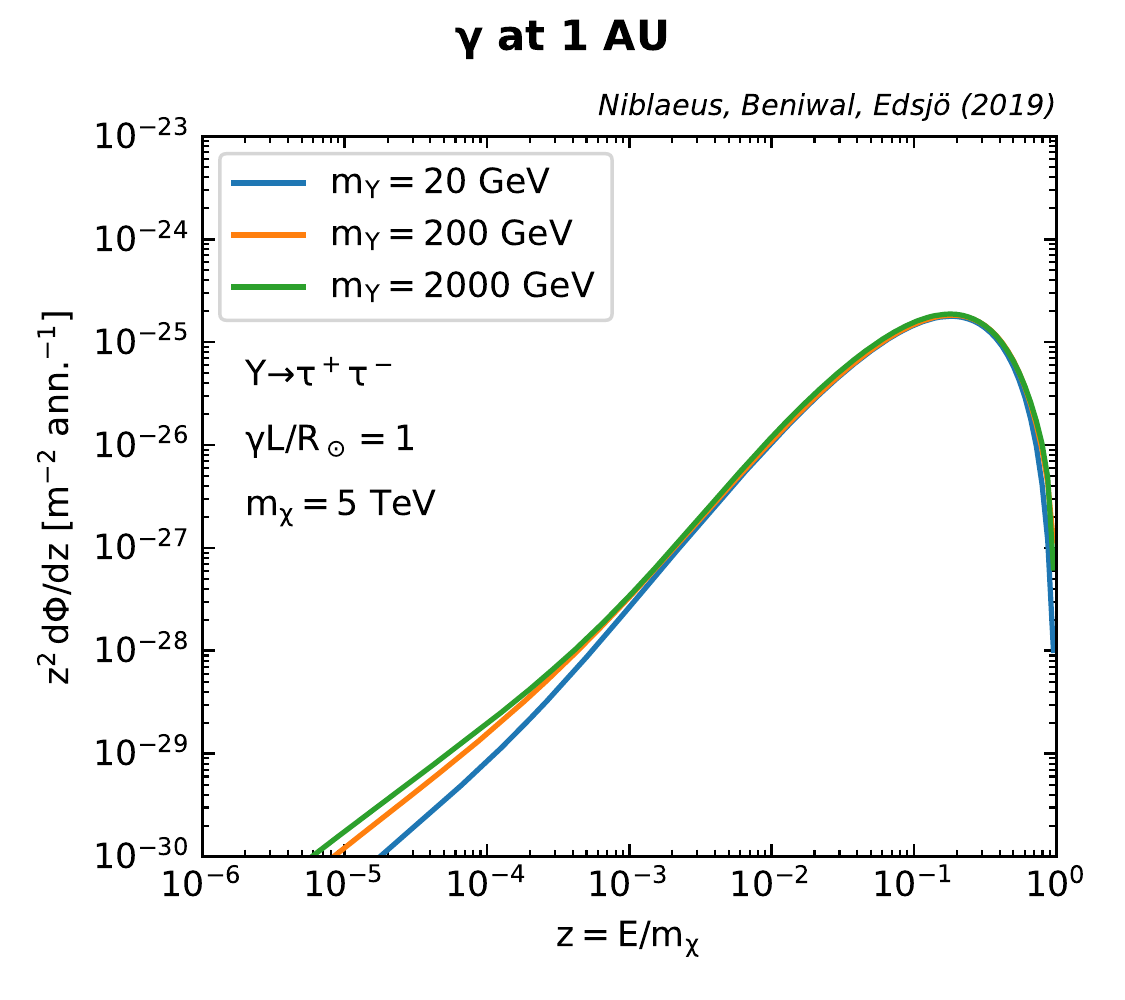}
    \caption{Same as figure~\ref{fig:nu-flux-medmass} for gamma ray fluxes.} 
    \label{fig:gamma-flux-medmass}
\end{figure}

In figures~\ref{fig:nu-flux-medmass} and~\ref{fig:gamma-flux-medmass}, we show the effect of varying the mediator mass ($m_Y$) on the muon neutrino and gamma ray fluxes at 1 AU. We can see that in accordance with the conclusions in ref.~\cite{Leane:2017vag}, the fluxes are not very sensitive to the value of the mediator mass. However, there are some differences especially in the case of the $b\overline{b}$ decay channel, where the flux is softened for higher mediator mass, due to more particles being produced in the hadronisation.

In comparison with figure~\ref{fig:neutrino-changing_Ldec}, we can see in figure~\ref{fig:nu-flux-medmass} the effect of neutrino oscillations due to higher neutrino energies, especially in the right panel for the $\tau^+\tau^-$ decay channel. Apart from the oscillations due to $\lambda_{21}$, showing at low $z \equiv E/m_\chi$, we clearly see the impact of  oscillations due to $\lambda_{3i}$ for $z\gtrsim 0.4$.

We also emphasise that the neutrino flux from a long-lived mediator decay is not suppressed by neutrino absorption in the Sun to the same extent as in the standard scenario. This absorption leads to an almost complete suppression of the neutrino flux above neutrino energies of $\sim\SI{1}{\tev}$ in the standard scenario, regardless of the value of $m_\chi$, whereas in the long-lived mediator scenario there is a significant flux also at higher energies, as evident in figure~\ref{fig:nu-flux-medmass}. Here the flux extends up to several \si{\tev} in neutrino energy. This leads to a significant increase in the expected event rate in neutrino telescopes and much more sensitivity to larger $m_\chi$, due to the fact that these telescopes are most sensitive to high energy neutrinos.

\subsection{Effect of a full three-dimensional decay treatment}\label{sec:results_3d}
\begin{figure}[t]
    \centering
    \includegraphics[width=0.49\textwidth]{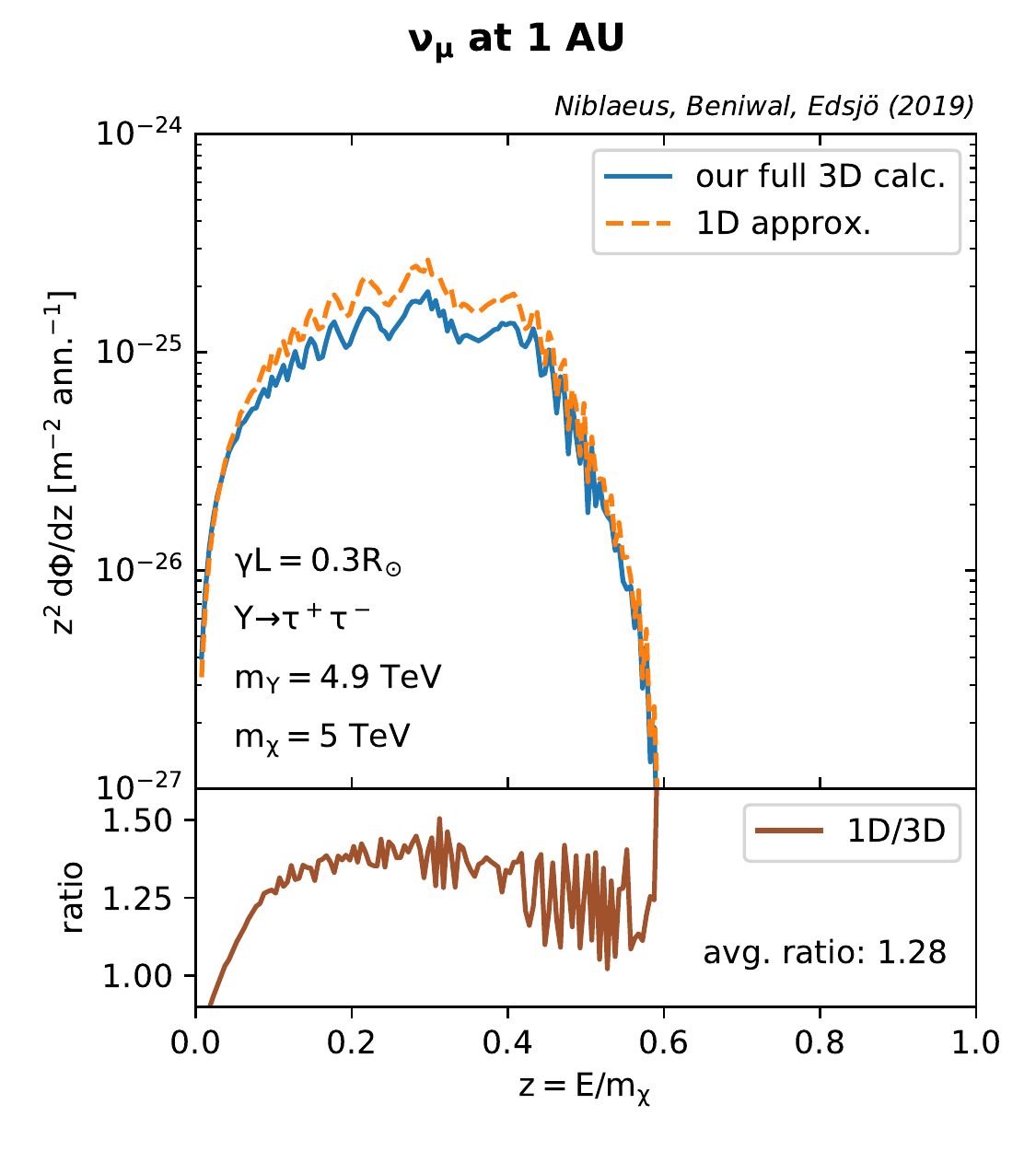}
    \caption{Muon neutrino fluxes at 1 AU from $Y \rightarrow \tau^+ \tau^-$ for $\gamma L/R_\odot=0.3$, $m_\chi=\SI{5000}{\gev}$ and $m_Y=\SI{4900}{\gev}$.~This parameter combination is chosen to be sensitive to our full three-dimensional calculation of the direction of the mediator decay products (as compared to the one-dimensional approximation where the decay products follow the mediator trajectory). Our choice for the simulation parameters are summarised in table~\ref{tab:sim_setup} (Scenario C).}
    \label{fig:3dvs1d}
\end{figure}

We use the full three-dimensional direction in all our simulations.~As we do not save the individual positions for gamma rays and charged particles from the decays, this effect is relevant only for neutrinos, where we follow the trajectory of each individual neutrino.~In figure~\ref{fig:3dvs1d}, we show the impact of a full three-dimensional calculation relative to the one-dimensional approximation where the mediator decay products follow the same direction as the mediator.

We can see that for the chosen parameter combination, the difference can be rather large, with the ratio reaching $\sim 1.5$ at some points and with an average ratio of $1.28$. However, we stress that this is a special case shown to highlight this effect, and in most cases, the effect will be much smaller. Thus, although we use the full three-dimensional calculation in all our simulations, the one-dimensional approximation will often be acceptable.

\subsection{Complementarity between gamma ray and neutrino searches}\label{sec:results_nuvsga}

\begin{figure}[t]
    \centering
    \includegraphics[width=0.49\textwidth]{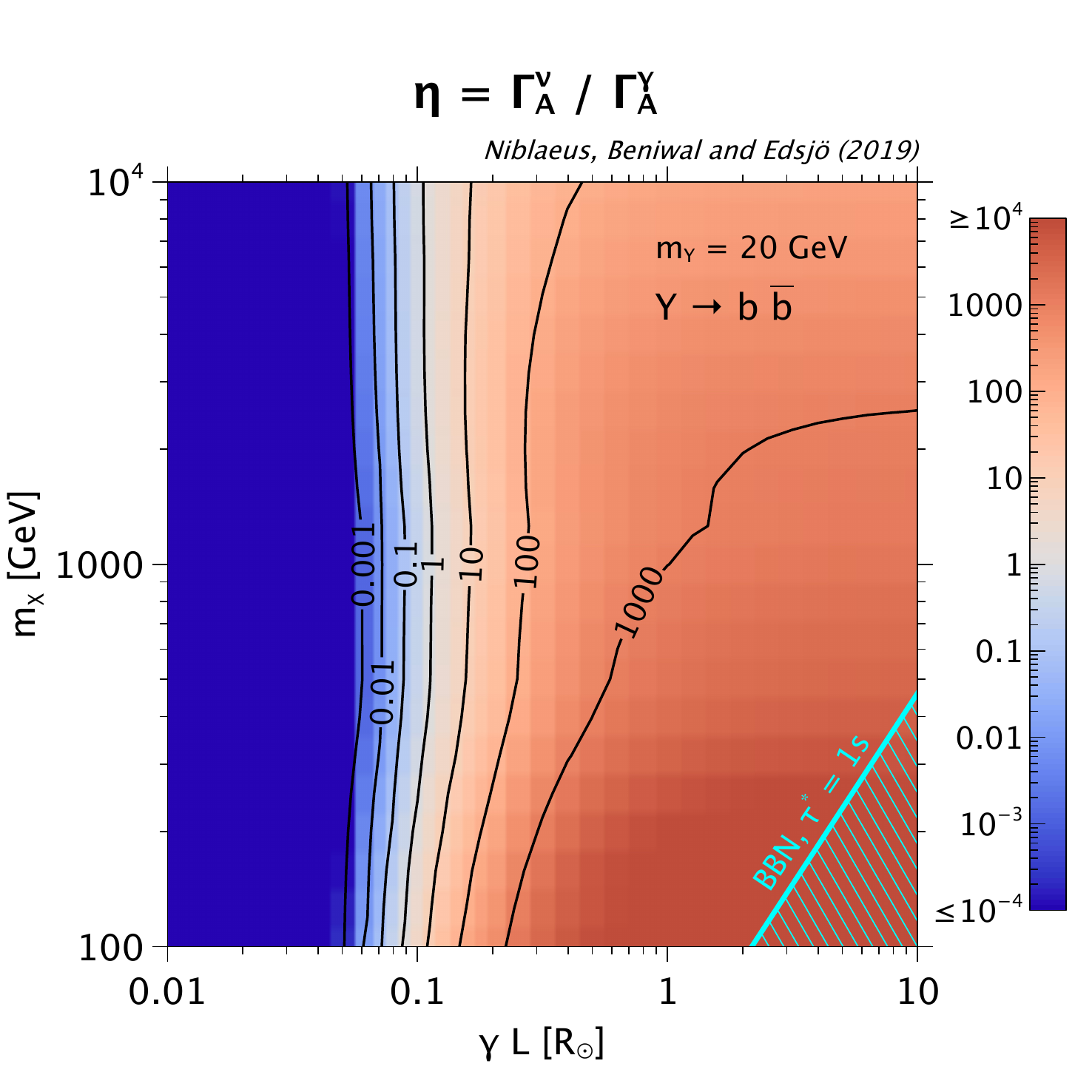}
    \includegraphics[width=0.49\textwidth]{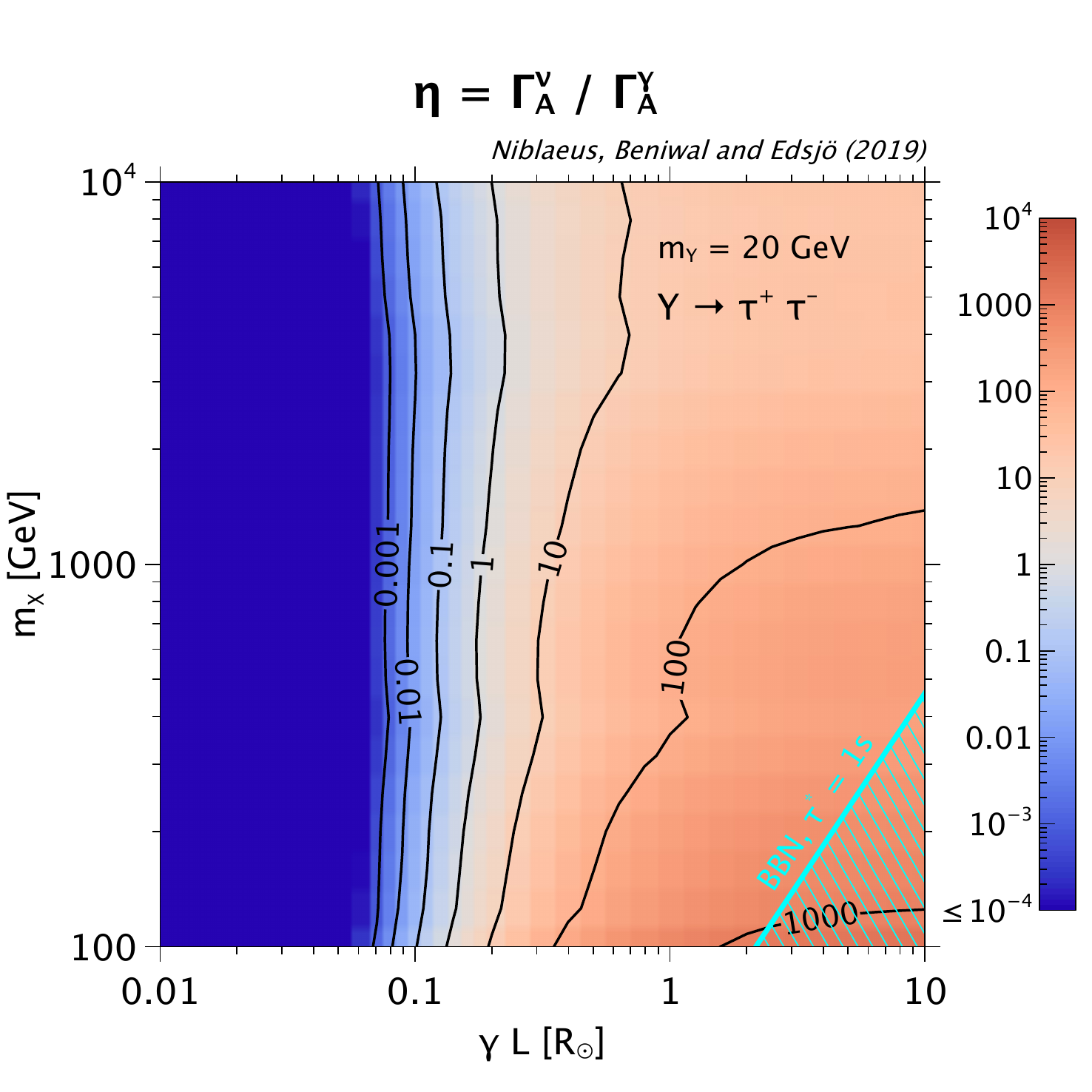}
    \caption{The ratio $\eta = \Gamma_A^\nu/\Gamma_A^\gamma$ in the ($\gamma L,\;m_\chi$) plane for $Y \rightarrow b\ovr{b}$ (\emph{left panel}) and $Y \rightarrow \tau^+ \tau^-$ (\emph{right panel}). In both panels, the mediator mass is $m_Y=20$ GeV. We also show the limit from BBN, eq.~\eqref{eq:lim_BBN}, for $\tau^*=1$\,s. 
    Note that  in the left panel of these figures, $\eta$ is much smaller than $10^{-4}$, while in the lower right of the left panel, $\eta$ is larger than $10^4$. We only include the range $\eta \in [10^{-4},10^4]$ in the color scale to show the differences in the most interesting region more clearly.
    }
    \label{fig:2dmed20}
\end{figure}

In figure~\ref{fig:2dmed20}, we show the ratio $\eta$ as defined in eq.~\eqref{eqn:ratio_eta} in the ($\gamma L,\;m_\chi$) plane for $Y \rightarrow b\ovr{b}$ (\emph{left panel}) and $Y \rightarrow \tau^+ \tau^-$ (\emph{right panel}) with $m_Y=20$ GeV\@. We also show the BBN limit for $\tau^*=1$\;s from eq.~\eqref{eq:lim_BBN}. From these figures, we can draw a few conclusions: 
\begin{itemize}
    \item The gamma ray limits are stronger at higher $\gamma L$ values. This agrees with our expectation as more mediators decay outside of the Sun give more gamma rays;
    \item The gamma ray limits are stronger for the $b\ovr{b}$ channel than the $\tau^+ \tau^-$ channel as we get proportionally more gamma rays than neutrinos for the $b\ovr{b}$ channel;
    \item The transition from gamma ray to neutrino domination is relatively deep inside the Sun at $\gamma L \simeq 0.1\,R_\odot$. The reason that we can get so many gamma rays even for such small $\gamma L$ values is that the tail of the decay probability distribution for the mediator outside the Sun is still large enough;
    \item The transition between neutrino and gamma ray domination is not very sensitive to the DM mass $m_\chi$ nor the decay channel. As we have seen before, we also do not expect a strong dependence on the mediator mass $m_Y$;
    \item The BBN constraint (shown here for $\tau^*=1$\,s) is not ruling out large parts of this parameter space.
\end{itemize}

\begin{figure}[t]
    \centering
    \includegraphics[width=0.49\textwidth]{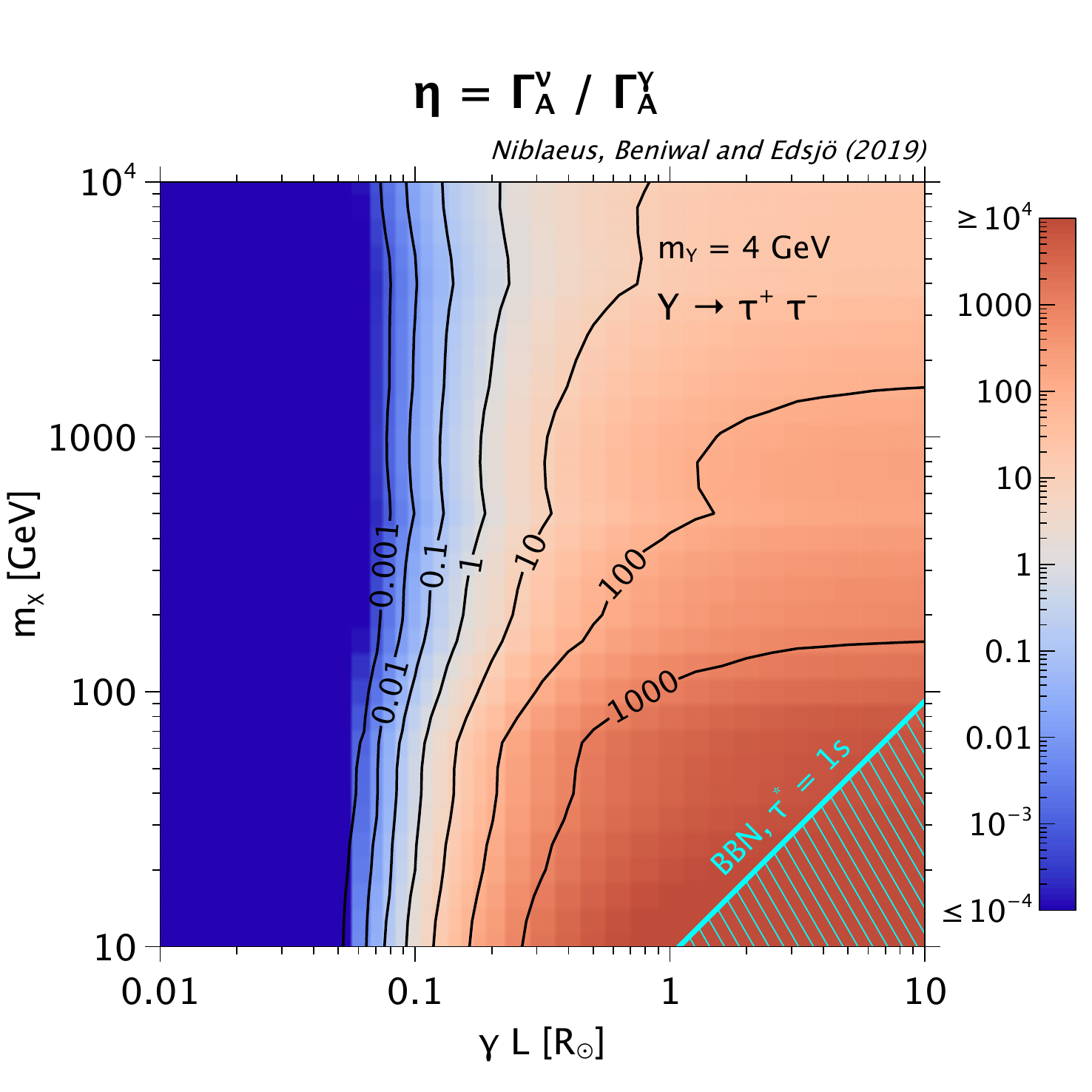}
    \caption{Same as figure~\ref{fig:2dmed20} but with $m_Y=\SI{4}{\gev}$ and only for $Y\to\tau^+\tau^-$.
    }
    \label{fig:2dmed4}
\end{figure}

For the sake of completeness, we show in figure~\ref{fig:2dmed4} the results for $Y \rightarrow \tau^+ \tau^-$ with $m_Y=4$ GeV (note that the vertical scale here extends to an order of magnitude lower DM masses than in figure~\ref{fig:2dmed20}).~As expected, the results for $m_Y=4$ GeV are very similar to those for $m_Y=20$ GeV\@.~The BBN limit is less constraining in this case though, as expected from eq.~\eqref{eq:lim_BBN}. Note that near the lower edge of figure~\ref{fig:2dmed4}, the boost of the mediator is relatively low which means that the signal will be less point-like. We do not include the extension of the source in the limit calculations, but both gamma ray and neutrino limits are affected in a similar way and to some extent this cancels in the ratio.

As can be seen in figures~\ref{fig:2dmed20} and \ref{fig:2dmed4}, neutrino searches are more constraining in the left regions (low $\gamma L$ values).~This is easy to understand as the probability for the mediator to decay outside of the Sun drops significantly for low $\gamma L$ values, whereas the neutrino fluxes are less affected. In fact, $\eta$ can be arbitrarily small in the left part of these figures.

\begin{figure}[t]
    \centering
    \includegraphics[width=0.32\textwidth]{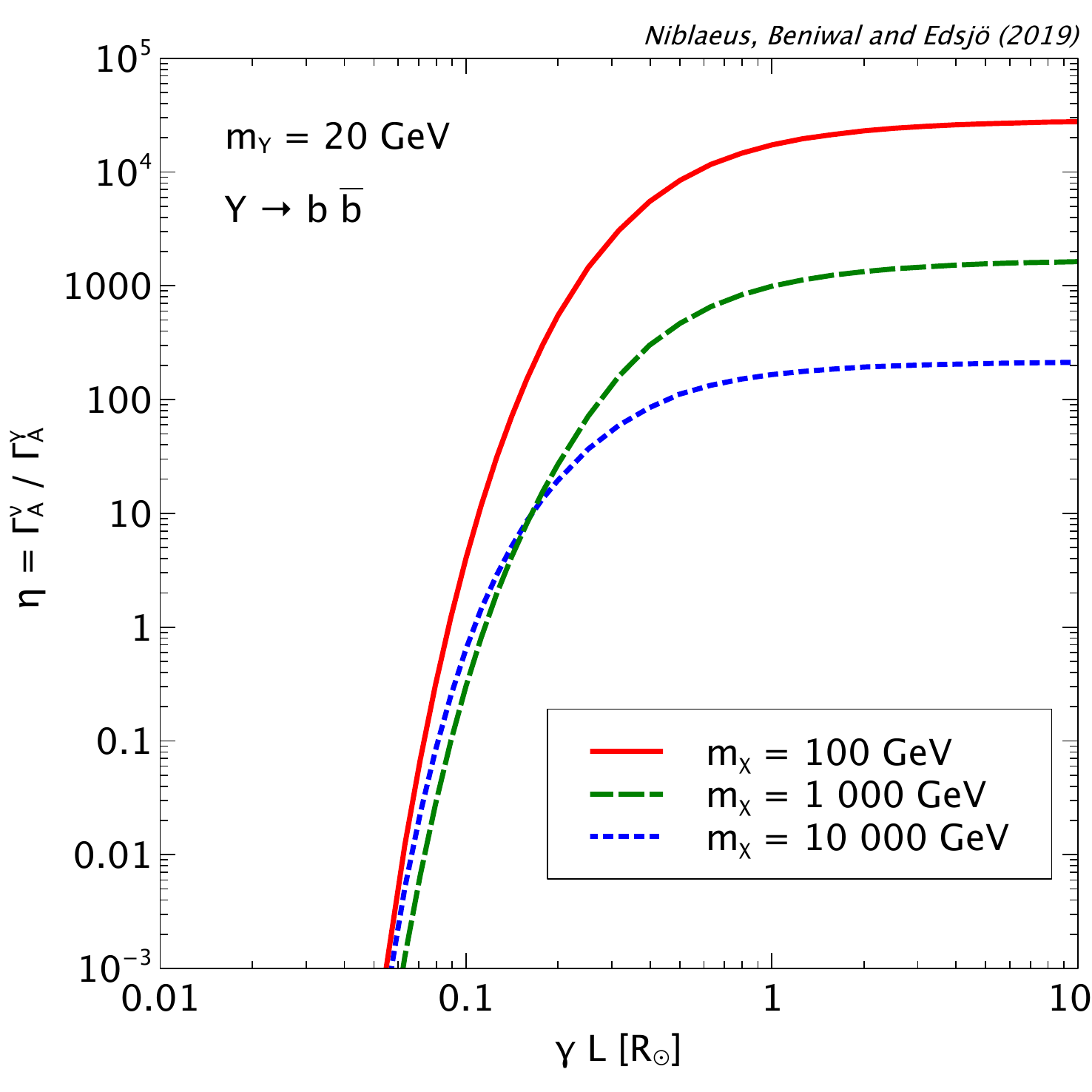}
    \includegraphics[width=0.32\textwidth]{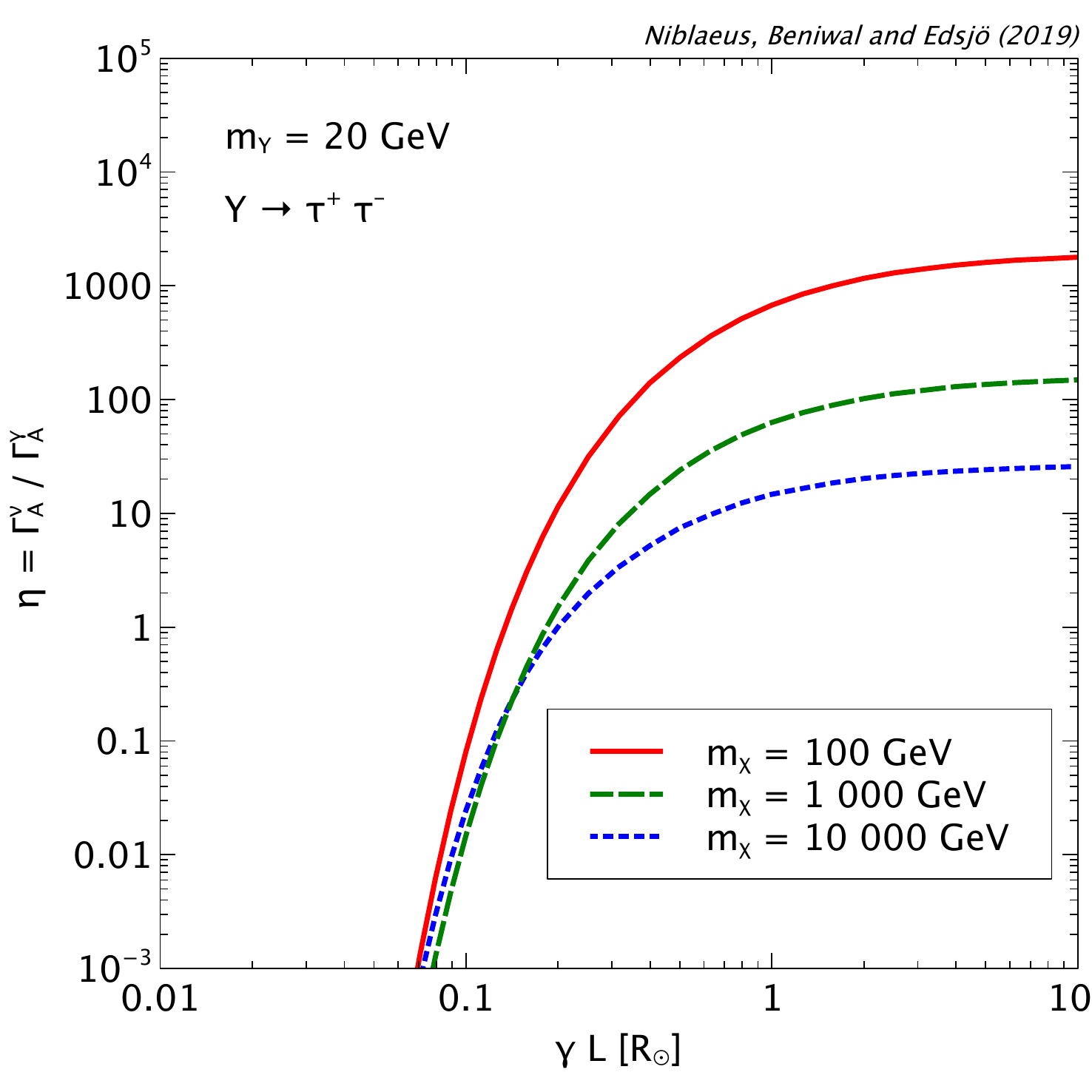}
    \includegraphics[width=0.32\textwidth]{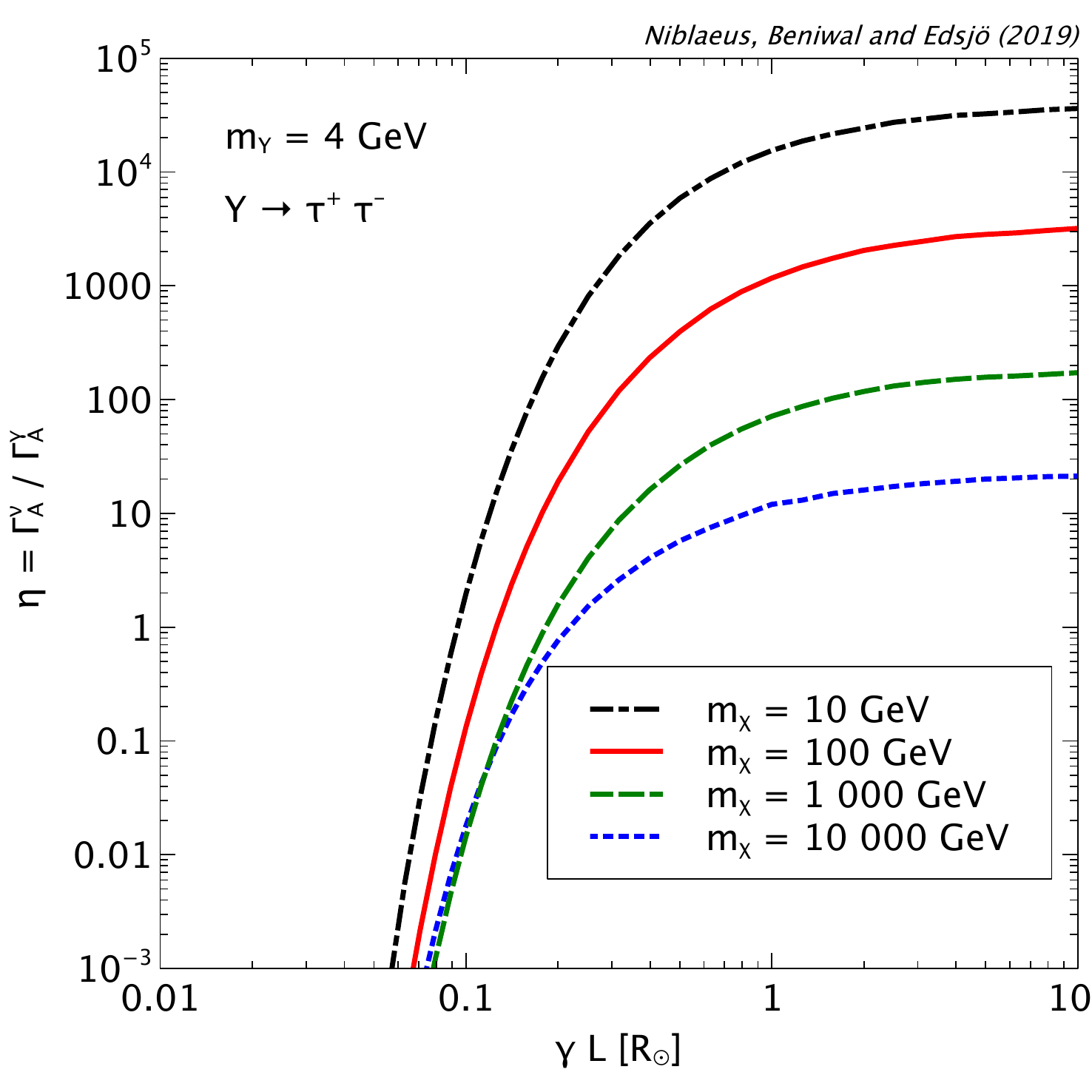}
    \caption{The ratio $\eta = \Gamma_A^\nu/\Gamma_A^\gamma$ versus $\gamma L$ for $m_Y=20$ GeV, $Y \rightarrow b \ovr{b}$ (\emph{left panel}), $m_Y=20$ GeV, $Y \rightarrow \tau^+ \tau^-$ (\emph{centre panel}) and $m_Y=4$ GeV, $Y \rightarrow \tau^+ \tau^-$ (\emph{right panel}).}
    \label{fig:ratio-vs-gl}
\end{figure}

\begin{figure}[t]
    \centering
    \includegraphics[width=0.32\textwidth]{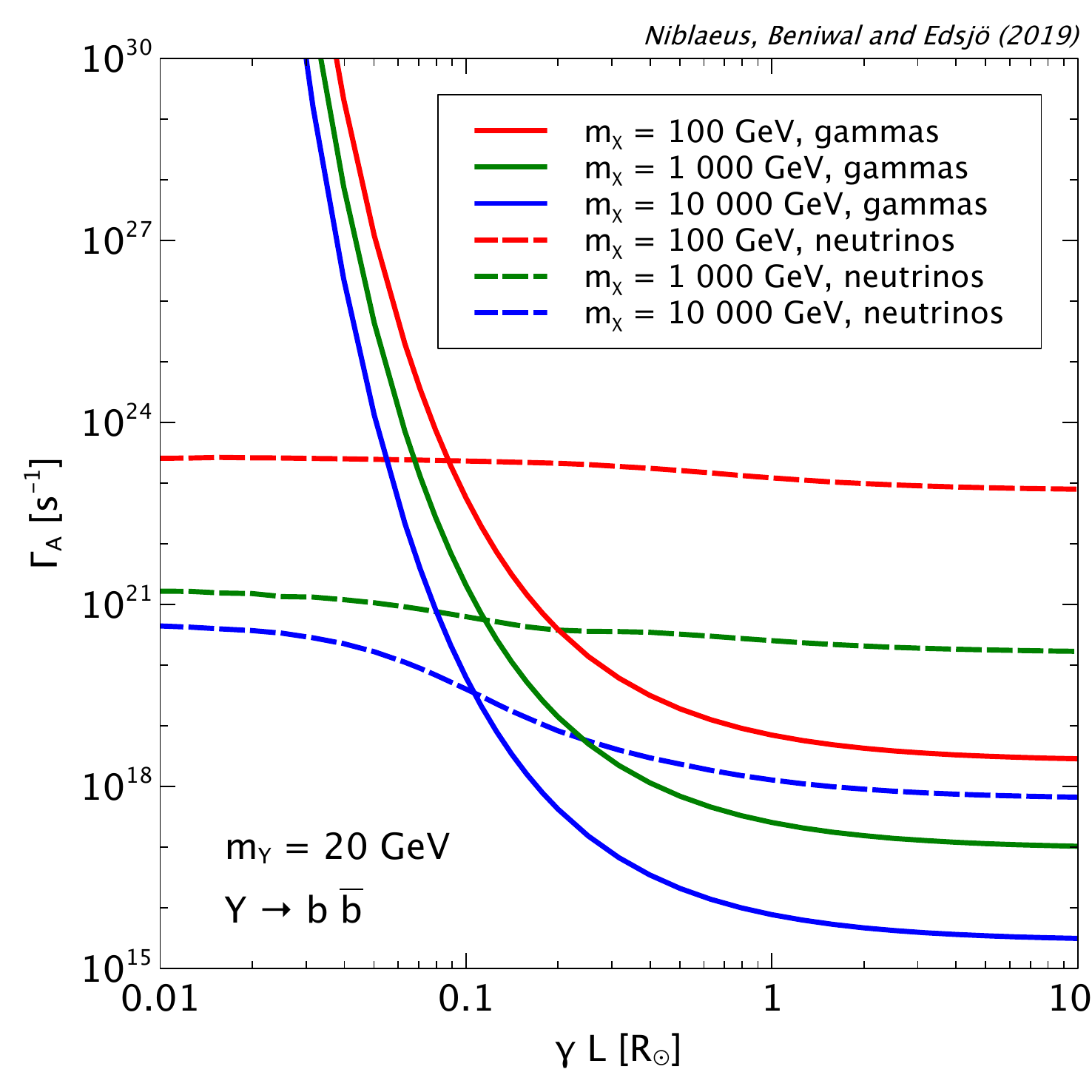}
    \includegraphics[width=0.32\textwidth]{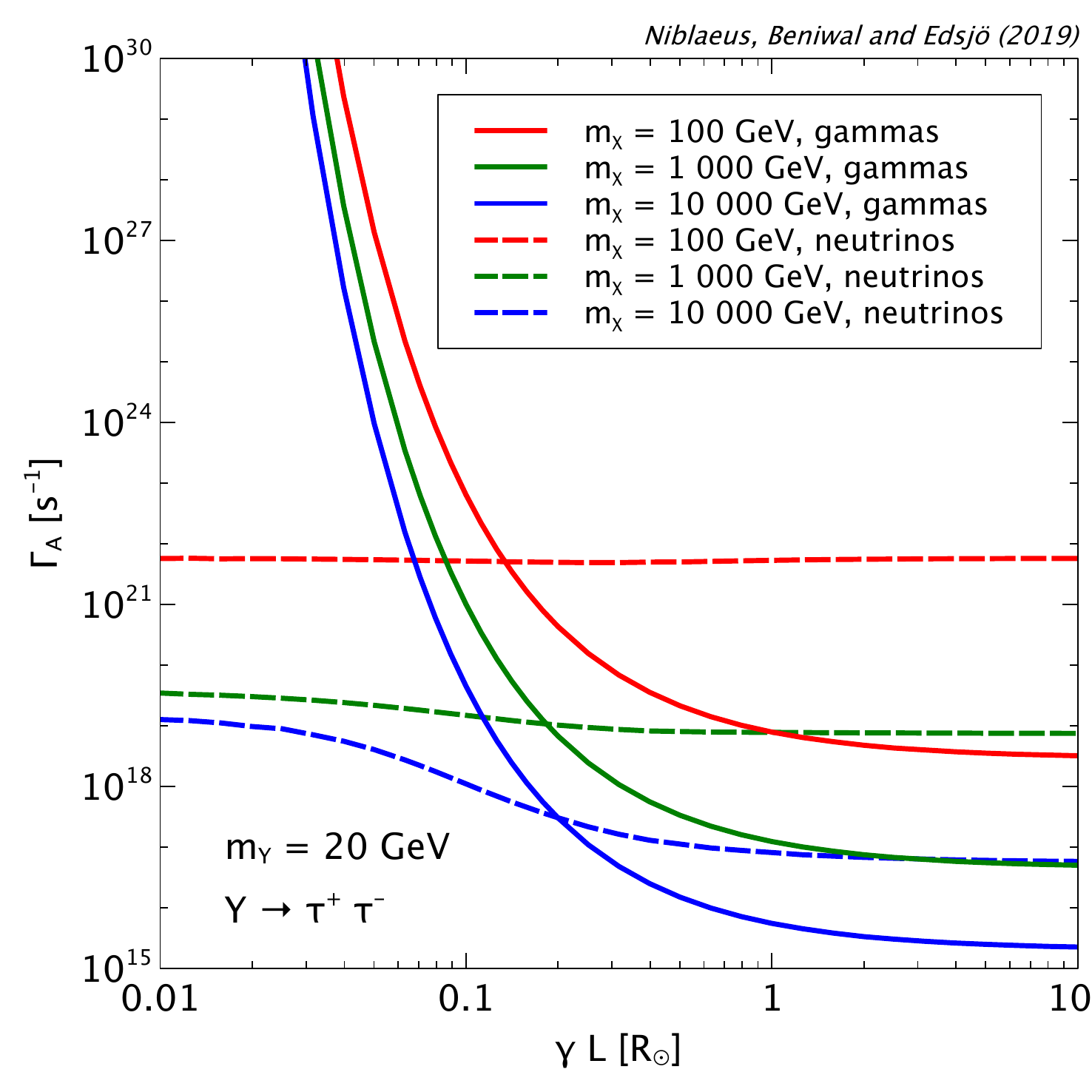}
    \includegraphics[width=0.32\textwidth]{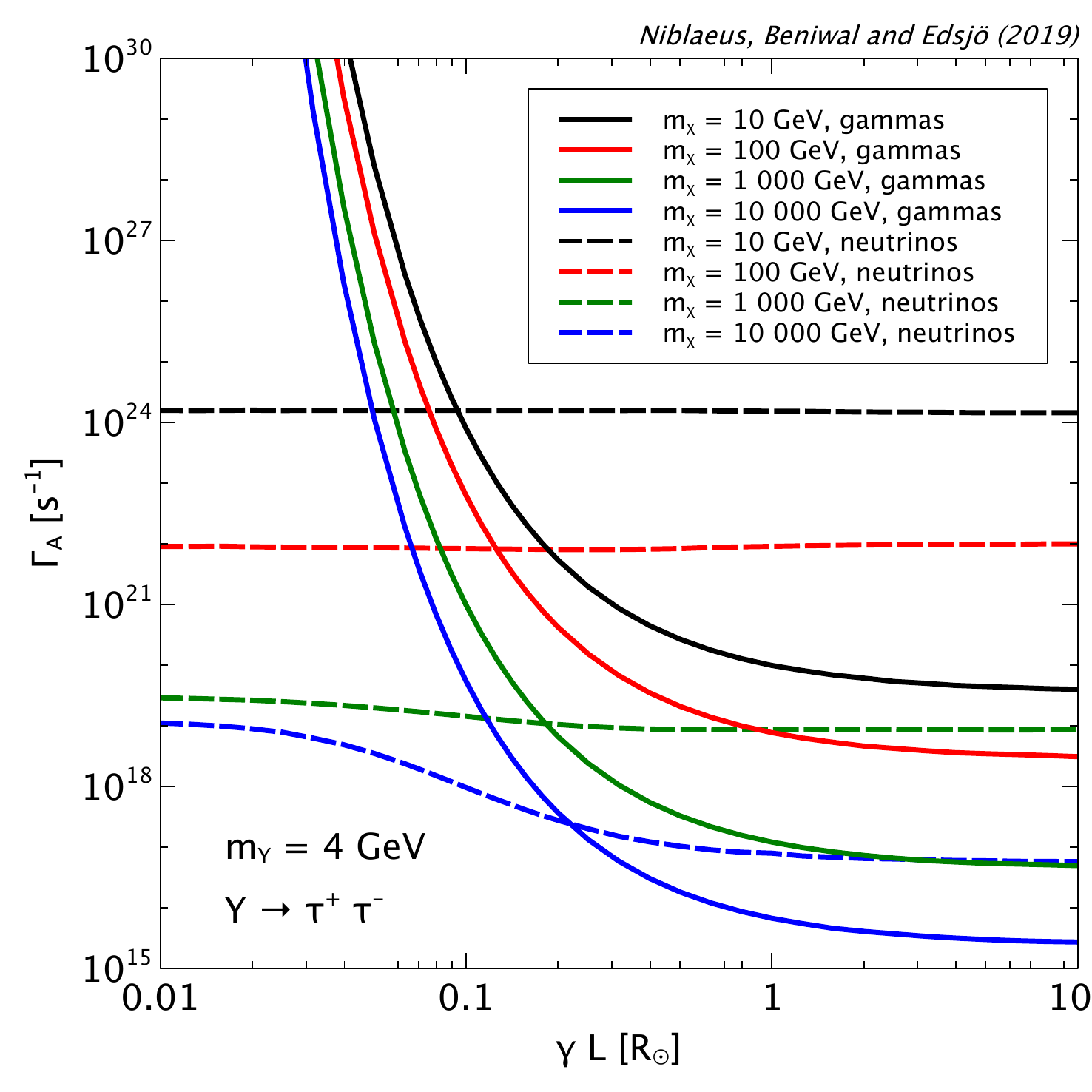}
    \caption{Upper limits on the DM annihilation rate, $\Gamma_A$ from neutrino and gamma ray searches for $m_Y=20$ GeV, $Y \rightarrow b \ovr{b}$ (\emph{left panel}), $m_Y=20$ GeV, $Y \rightarrow \tau^+ \tau^-$ (\emph{centre panel}) and $m_Y=4$ GeV, $Y \rightarrow \tau^+ \tau^-$ (\emph{right panel}). The region above the curves is excluded.}
    \label{fig:ga-vs-gl}
\end{figure}

To explore this in more detail, in figure~\ref{fig:ratio-vs-gl}, we show the ratio $\eta$ versus $\gamma L$ for a few different DM masses $m_\chi$ and the cases shown in the previous figures.~It is evident that the transition between gamma ray domination (to the right in the figures) and neutrino domination (to the left in the figures) is very sharp. We also see that for high $\gamma L$, the ratio $\eta$ is significantly larger for small DM masses $m_\chi$. This makes sense as the sensitivity of neutrino telescopes rises approximately as $E_\nu^2$ where $E_\nu$ is the neutrino energy.~The typical neutrino energies will be proportional to $m_\chi$ and hence we expect proportionally stronger limits from neutrino telescopes for higher masses, which is what we see. In spite of this though, the transition from gamma ray to neutrino domination happens at roughly the same $\gamma L \simeq 0.1\,R_\odot$. 

To get a more clear picture of the actual numbers and strengths of the signals involved, we show in figure~\ref{fig:ga-vs-gl} the limits on the DM annihilation rate $\Gamma_A$ from neutrinos and gamma rays for our three different cases.~From this figure it is evident that the limit on $\Gamma_A$ from gamma rays depend strongly on $\gamma L$. This is of course expected as essentially we are seeing the inverse of the probability that the mediator decays outside of the Sun. On the other hand, the limits on $\Gamma_A$ from neutrinos does not show an equally strong dependence on $\gamma L$.~We can understand this from the fact that the main effect we expect for neutrinos is that for large DM masses $m_\chi$, we get significantly less absorption for larger $\gamma L$ values since the mediator (in this case) decays away from the very dense region in the solar centre. In figure~\ref{fig:ga-vs-gl}, this is clearly seen as the neutrino curves give significantly better $\Gamma_A$ limits at high $\gamma L$ values for large DM masses. The difference in the limits at high and low $\gamma L$ values is more than two orders of magnitude for a DM mass of $m_\chi=10^4$\;GeV.

\begin{figure}[t]
    \centering
    \includegraphics[width=0.49\textwidth]{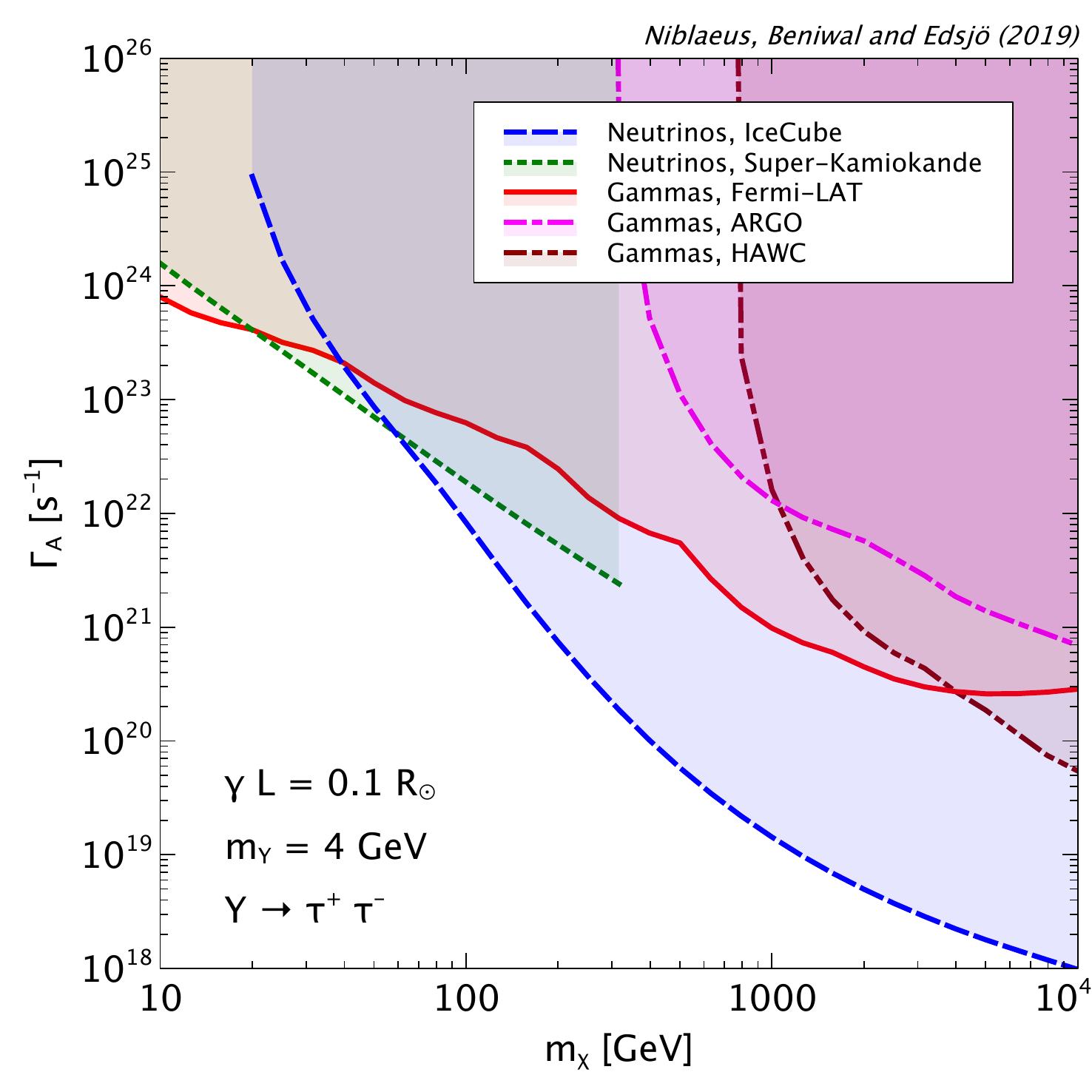}
    \caption{Upper limits on the DM annihilation rate, $\Gamma_A$ versus the DM mass $m_\chi$ from neutrino and gamma ray searches respectively for $m_Y=4$ GeV, $Y \rightarrow \tau^+ \tau^-$ and $\gamma L=0.1\,R_\odot$. Note that if we increase (decrease) $\gamma L$, the most significant effect in this figure is that the $\Gamma_A$ curves for gamma rays move down (up). The region above the curves is excluded.}
    \label{fig:ga-vs-mx}
\end{figure}

In figure~\ref{fig:ga-vs-mx}, we finally show the limit on the DM annihilation rate $\Gamma_A$ versus the DM mass $m_\chi$ for $Y \rightarrow \tau^+ \tau^-$ and $m_Y=4$ GeV. In this figure, we have set $\gamma L=0.1\,R_\odot$, i.e., close to the transition region between gamma ray and neutrino domination. We can clearly see that at low DM masses, the gamma ray limits are the strongest, and for higher masses we get stronger limits from neutrino telescopes, first from Super-Kamiokande and then from IceCube. We also see that the most constraining gamma ray experiment for the DM masses we have considered is \fermi, except at the very high DM masses, where HAWC starts being more constraining. Note that if we increase (decrease) $\gamma L$, the most significant effect in this figure is that the $\Gamma_A$ curves for gamma rays move down (up). If we decrease $\gamma L$ to very low values, the neutrino curves move up for the highest masses (due to absorption effects). This means that in most of the parameter space in figures~\ref{fig:2dmed20} and \ref{fig:2dmed4}, the strongest limits mostly come from IceCube for neutrinos and \fermi\ for gamma rays.~The exceptions are the lowest DM masses where the Super-Kamiokande limits are more constraining than IceCube for neutrinos, and the highest masses where HAWC is more constraining than \fermi\ for gamma rays.

For neutrino telescopes, we expect some progress in the coming years.~The IceCube upgrade \cite{Kowalski:2019yxj} is expected to improve the sensitivity somewhat for the lower masses. When IceCube-Gen2 \cite{Ackermann:2017pja,Kowalski:2019yxj} (a 10 km$^3$ volume neutrino telescope) is completed, it is expected to increase the sensitivity by a factor of a few. In the Mediterranean, KM3NeT \cite{Adrian-Martinez:2016fdl} is undergoing construction, with two parts of the detector: ORCA for lower energies, and ARCA for higher energies. The main benefit of these detectors is the longer scattering length in Mediterranean water compared to ice, thus giving better angular resolution.~The net effect of this is to improve the limits by a factor of a few compared to current limits. Hence, we do expect improvements of factors of a few in the coming years. However, our general results on where the gamma ray or neutrino searches are most sensitive will not change significantly.

Regarding gamma ray telescopes, we also expect an increased sensitivity with the upcoming LHAASO experiment \cite{Bai:2019khm, Cao:2010zz}. Especially at higher energies, this will be several times more sensitive to gamma rays than current experiments.

\subsection{The muon channel}\label{sec:results_muon}
In refs.~\cite{Adrian-Martinez:2016ujo,Ardid:2017lry}, the authors investigate limits on secluded DM models from ANTARES and IceCube in a similar setup as we do here.~However, they focus on mediator decay channels where neutrinos are expected to dominate.~For instance, in ref.~\cite{Ardid:2017lry}, the authors focus on direct decay to neutrinos, decay to $\pi^+ \pi^-$ and decay to $\mu^+ \mu^-$.~To compare with the latter study, we will focus on the muon channel, i.e.,\ $Y \to \mu^+ \mu^-$.

For the muonic decay channel, one can also search for the muon tracks from mediators decaying in or near the detector, rather than searching for tracks induced by neutrino scattering events. We do not consider this possibility here but note that in some parts of the parameter space, this type of search results in slightly stronger limits than neutrino searches from muon decays \cite{Adrian-Martinez:2016ujo}.

\begin{figure}[t] 
    \centering
    \includegraphics[width=0.49\textwidth]{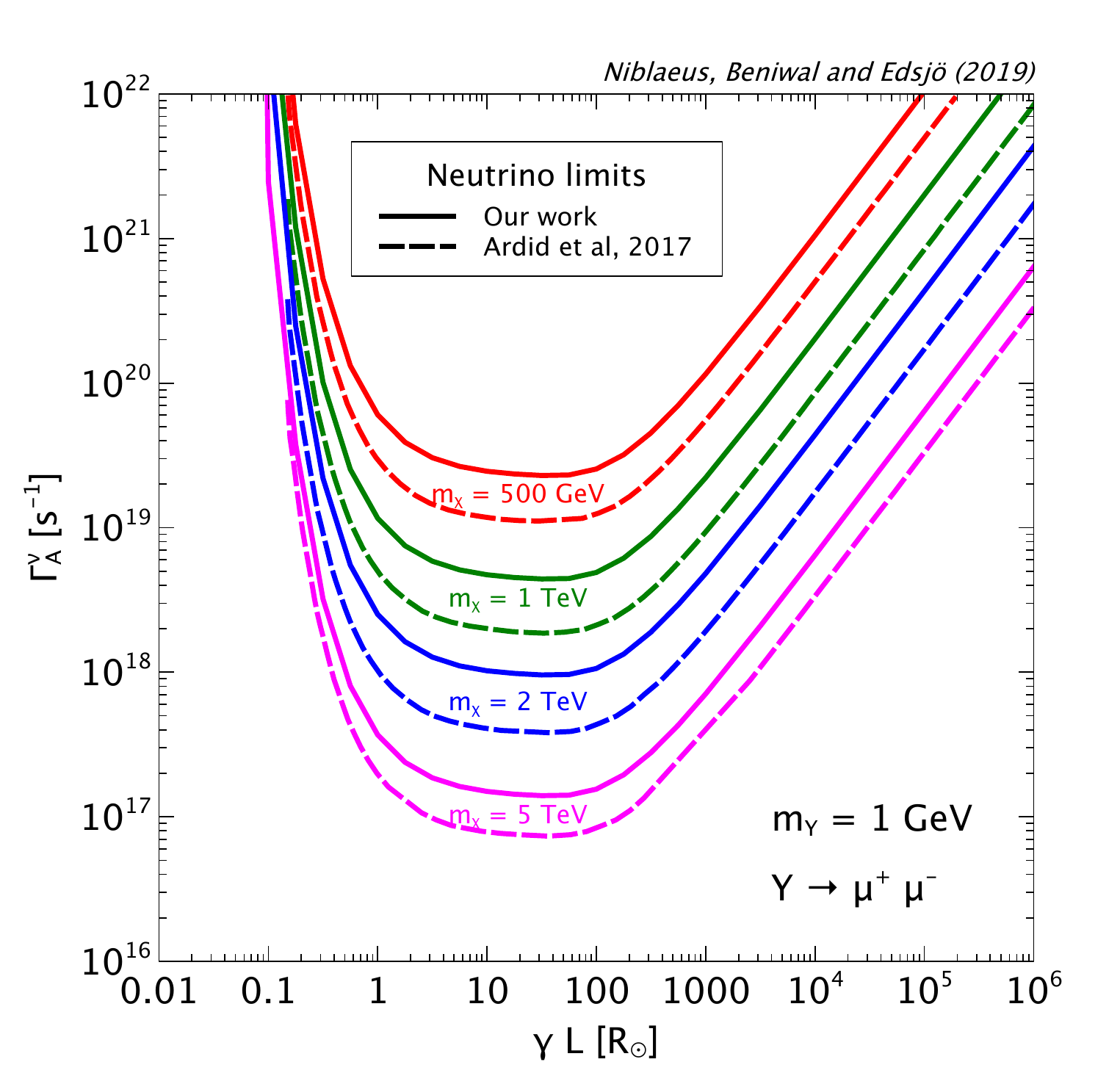}
    \vspace{-1.5em}
    \includegraphics[width=0.49\textwidth]{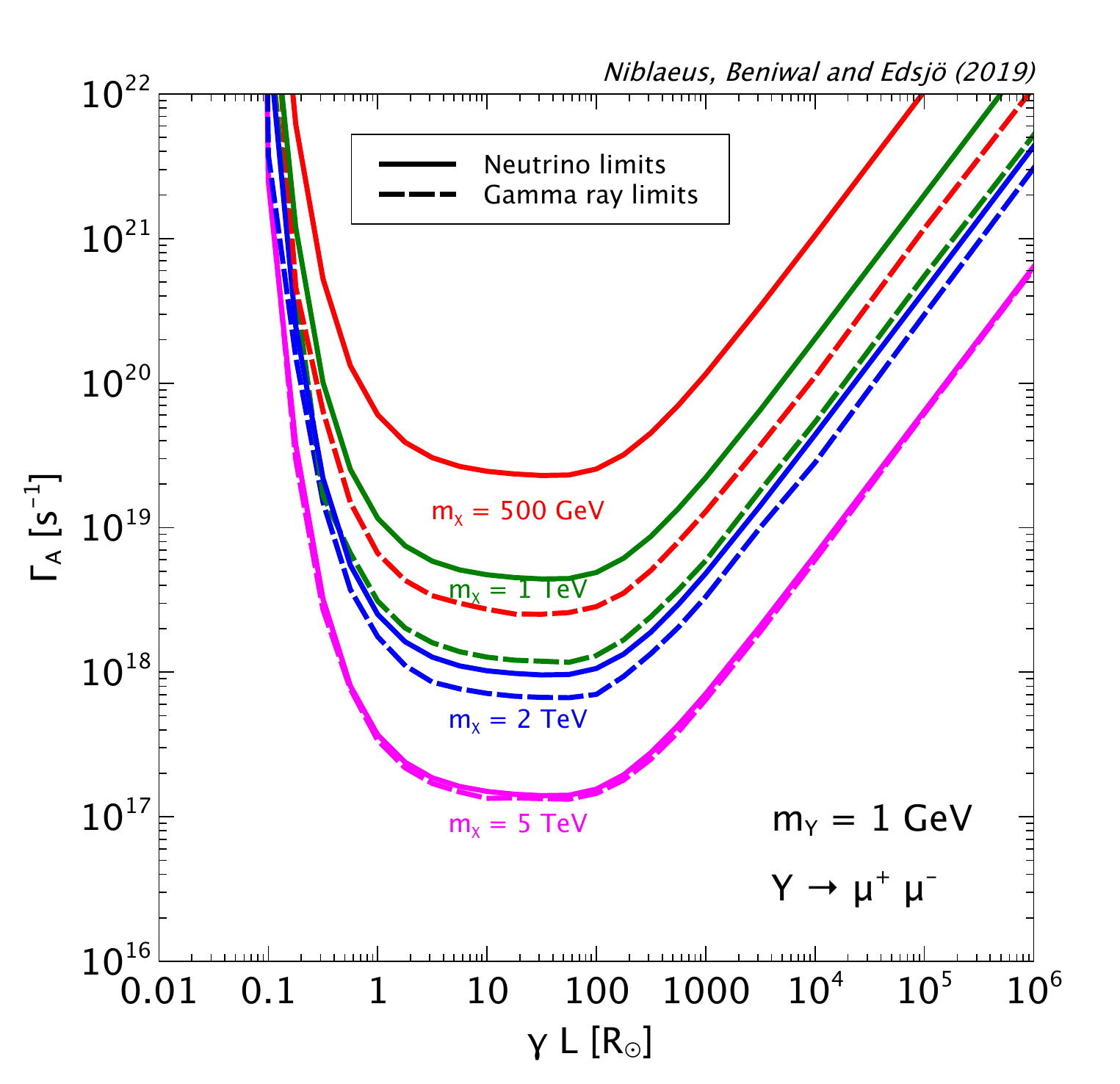} \vspace{1em}
    \caption{Upper limits on the DM annihilation rate, $\Gamma_A$ versus the boosted decay length $\gamma L$ for $m_Y=1$ GeV, $Y \rightarrow \mu^+ \mu^-$. In the left panel, we show our results as solid curves and the results from ref.~\cite{Ardid:2017lry} as dashed curves. In the right panel, we again show our results as solid curves, but compare with the limits coming from gamma ray searches as dashed curves. The region above the curves is excluded.}
    \label{fig:icmu-comp}
\end{figure}

In the left panel of figure~\ref{fig:icmu-comp}, we show the limit on the annihilation rate, $\Gamma_A^\nu$ versus the boosted decay length $\gamma L$. Note that we obtain qualitatively very similar results as in ref.~\cite{Ardid:2017lry}. However, our limits are a factor of 2--3 worse.~It is hard to pinpoint exactly where this difference comes from. However, we note that we use quite different methods for estimating the limits. For IceCube, we use the limits from ref.~\cite{Aartsen:2016zhm} which are derived for a WIMP scenario using 3 years of data. We adapt it for our (similar) scenario as outlined in Appendix \ref{app:nulimits}. Thus, we expect our limits to be a reasonable approximation to the limits derived in ref.~\cite{Aartsen:2016zhm}.  In ref.~\cite{Ardid:2017lry} however, the authors derive their own limits from the publicly available (1 year) data for IceCube-79. One difference that we note is that in ref.~\cite{Ardid:2017lry}, they use a rather tight angular cut (their table~1) to define their signal region, whereas the IceCube analysis that we rely on in ref.~\cite{Aartsen:2016zhm} indicate a much larger angular resolution (their figure 4, right panel), which would indicate a larger signal region. A larger signal region also means a larger background region, which results in weaker limits.

On the other hand, we want to point out that even in case of mediator decay to muons, we do expect final state radiation off the muons.~In our \textsf{Pythia} simulations, this effect is included. In the right panel of figure~\ref{fig:icmu-comp}, we compare our neutrino limits with the gamma ray limits. Interestingly, we find that the gamma ray limits are stronger than the neutrino limits also in this scenario. This is particularly evident at smaller DM masses, where the difference is more than an order of magnitude. At higher masses, the limits are more comparable. Compared to figure~\ref{fig:ga-vs-gl} we here show results for much larger $\gamma L$ values (to compare with the results in ref.~\cite{Ardid:2017lry}). At high $\gamma L$ values, the increase in the $\Gamma_A$ limits is caused by a reduction in the flux at Earth due to the mediators decaying after 1 AU more often.

\begin{figure}[t]
    \centering
    \includegraphics[width=0.49\textwidth]{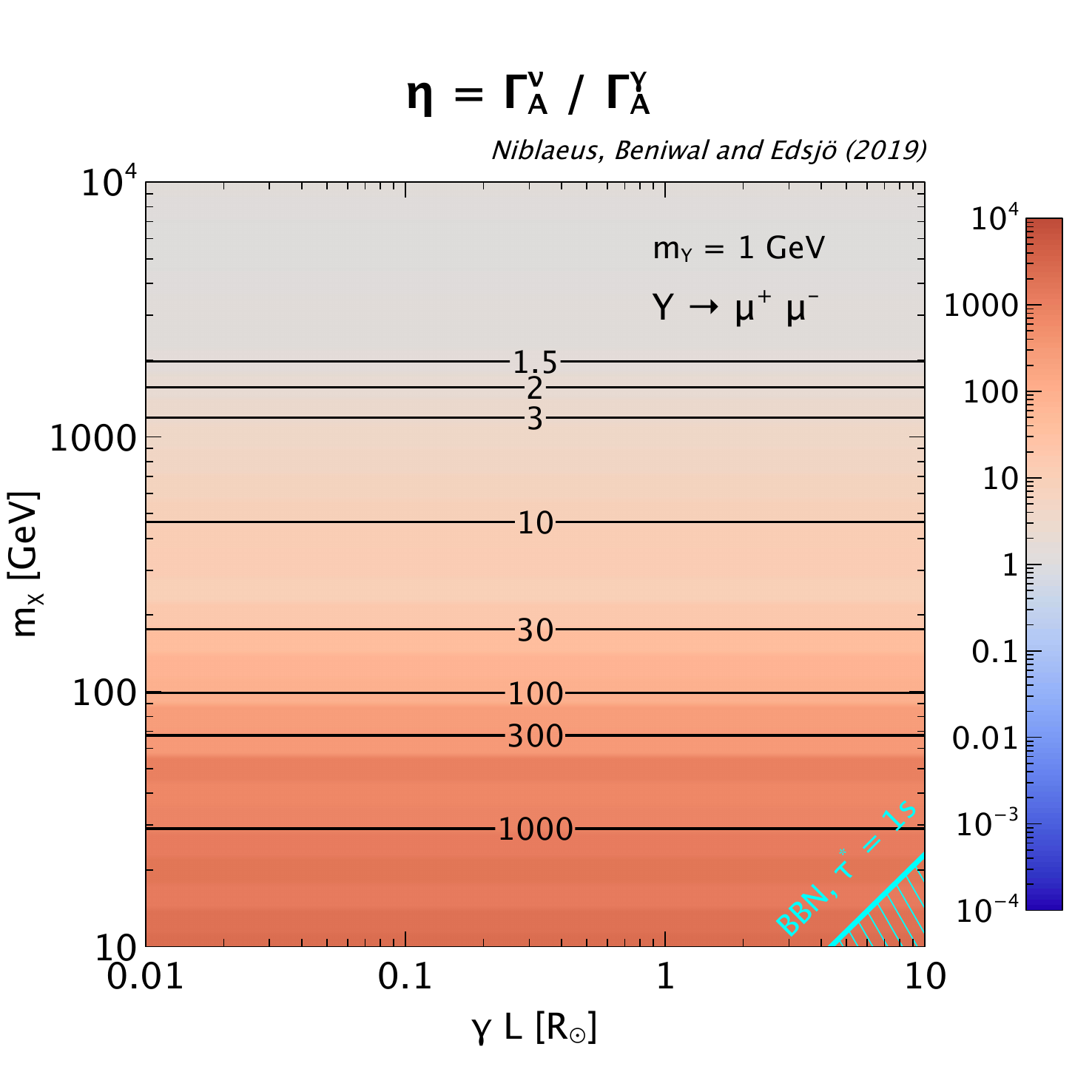}
    \caption{The ratio $\eta = \Gamma_A^\nu/\Gamma_A^\gamma$ in the ($\gamma L, m_\chi$) plane for $Y \rightarrow \mu^+ \mu^-$. In this figure, the mediator mass is $m_Y=1$ GeV. We also show the limit from BBN, eq.~\eqref{eq:lim_BBN}, for $\tau^*=1$ s.}
    \label{fig:icmu-2d}
\end{figure}

To relate with the previous subsection, we show the ratio $\eta=\Gamma_A^\nu / \Gamma_A^\gamma$ in the $(m_\chi,\gamma L)$ plane in figure~\ref{fig:icmu-2d}.~This figure looks quite different compared to figures~\ref{fig:2dmed20} and \ref{fig:2dmed4}. The reason is that in the muon channel, only mediator decays outside the Sun contribute and hence the neutrino and gamma ray fluxes scale with $\gamma L$ in the same way. When we lower $\gamma L$, we will get less muons decaying outside of the Sun, thus producing less gamma rays and neutrinos, whereas in figures~\ref{fig:2dmed20} and \ref{fig:2dmed4}, we get neutrinos also from the decays happening inside of the Sun.

In this case, especially for higher masses, the neutrino and gamma ray limits are comparable and improved sensitivities by e.g.,\ the IceCube Upgrade, IceCube-Gen2 \cite{Ackermann:2017pja,Kowalski:2019yxj} or KM3NeT \cite{Adrian-Martinez:2016fdl} may shift things in favour of neutrino telescopes. However, LHAASO \cite{Bai:2019khm,Cao:2010zz} will also offer improved sensitivities to the gamma ray flux at higher masses, so the net effect may not be that different.

\section{Conclusions}\label{sec:conclusions}
In this paper, we have considered a scenario where DM particles annihilate at the centre of the Sun into a pair of long-lived mediators. These mediators are emitted back-to-back and propagate away from  the annihilation point before decaying into a pair of stable SM particles. For a given value of the DM mass, mediator mass, mediator boosted decay length and decay channel, we predicted the energy spectra of neutrinos and gamma rays using our new version of the \wimpsim code. This code allows us to handle the DM annihilations in the solar centre, and propagation and decay of resulting mediators. It also stores each neutrino produced in the mediator decay and propagates it from the mediator decay point to the detector, and handles all neutrino interactions and oscillations including interactions with the detector material. For gamma rays and charged cosmic rays, the differential fluxes from mediator decays are collected and stored. 

In comparison with the standard scenario where DM particles annihilate in the centre of the Sun into short-lived SM particles that inject neutrinos close to the solar centre, we find that the long-lived mediator scenario leads to a harder neutrino spectra.~As the boosted decay length $\gamma L$ of the mediators is increased, the mediators decay on average further out from the centre.~Thus, the resulting neutrinos are subjected to much less interactions with the solar material before reaching the solar surface, thereby leading to a harder energy spectrum. This is of particular importance for neutrino telescopes as they are significantly more sensitive at higher neutrino energies.

For mediators that decay outside the Sun, other messenger particles of interest can be produced in the mediator decays and propagate to a detector on Earth. Here we have focused on gamma rays which are produced in the mediator decays from the decays of hadrons and in final state radiation from charged particles.

We find that the spectra of gamma rays and neutrinos mainly depends on the DM mass, the (boosted) decay length and decay channel of the mediator, whereas the mediator mass plays a sub-dominant role.~For instance, the DM mass determines the maximally possible neutrino or gamma ray energy, the decay length determines the impact of solar absorption on the energy spectrum for neutrinos, and the normalisation of the spectrum for gamma rays. On the other hand, the decay channel determines how the neutrinos and gamma rays are produced, and hence impacts the shape of the spectrum. 

To assess the strength of neutrino and gamma ray searches in the long-lived mediator scenario, we have compared our predictions against the neutrino observations from IceCube and Super-Kamiokande, and gamma ray observations from \fermi, HAWC and ARGO. We have also compared the limit that neutrino and gamma ray searches respectively can place on the DM annihilation rate $\Gamma_A$ for the long-lived mediator scenario, and looked at their ratio which determines the relative strength of the two searches. We found that gamma ray searches are most constraining for $\gamma L\gtrsim 0.1\,R_\odot$, while the sensitivity drops rapidly below this value in favour of neutrino telescopes. For the $\tau^+\tau^-$ decay channel of the mediator, which produces a harder neutrino spectrum, the neutrino telescopes are somewhat more sensitive at higher DM masses, also for larger $\gamma L$.~However, limits from gamma ray telescopes are still significantly stronger above $\gamma L\sim 0.1\,R_\odot$. We have performed our comparison between neutrino and gamma ray searches using approximations of the experimental limits. We want to emphasise though that our results depend very strongly on $\gamma L$ so even if the limits are improved by a factor of a few (with e.g.,\ a more accurate study or improved understanding of the gamma ray backgrounds), the general feature with a dividing line between the two searches at $\gamma L\simeq 0.1\,R_\odot$ is expected to remain.

Although gamma ray limits tend to be much more constraining for $\gamma L\gtrsim 0.1\,R_\odot$, there is still considerable strength in a multi-messenger approach, in particular, when it comes to the characterisation of the expected solar atmospheric background of neutrinos and gamma rays formed in cosmic ray cascades in the Sun. This mechanism is expected to produce both gamma rays and neutrinos with similar fluxes, whereas a signal from mediator decays could (depending on the decay channel) produce gamma ray and neutrino fluxes that differ to a much larger extent.~Thus, a multi-messenger approach is highly useful to disentangle any potential DM signal from known astrophysical backgrounds.

\acknowledgments{We thank K.~Hultqvist and M.~Rameez for helpful discussions. This work was supported by the Swedish Research Council (contract 621-2014-5772).~AB is supported by the Fund for Scientific Research F.N.R.S. through the F.6001.19 convention.}

\newpage
\appendix

\section{Neutrino telescope limits for the long-lived mediator scenario}\label{app:nulimits}
Limits on the neutrino flux from DM annihilations in the Sun are usually derived for specific DM models, typically involving a generic WIMP with a set of masses and annihilation channels (e.g., hard spectras like $\tau^+\tau^-$ or $W^+W^-$, or soft spectra like $b\ovr{b}$). The latest results derived in such a setting are the IceCube 3-year data \cite{Aartsen:2016zhm} and the Super-Kamiokande low-mass analysis \cite{Choi:2015ara}. Exceptions to this assumption is the IceCube 79-string data \cite{Aartsen:2016exj} where results are instead provided in a likelihood formalism applicable to any DM model.~Unfortunately, results like these are not available for the latest results from neither IceCube nor Super-Kamiokande.

\begin{figure}[t]
    \centering
    \includegraphics[width=0.49\textwidth]{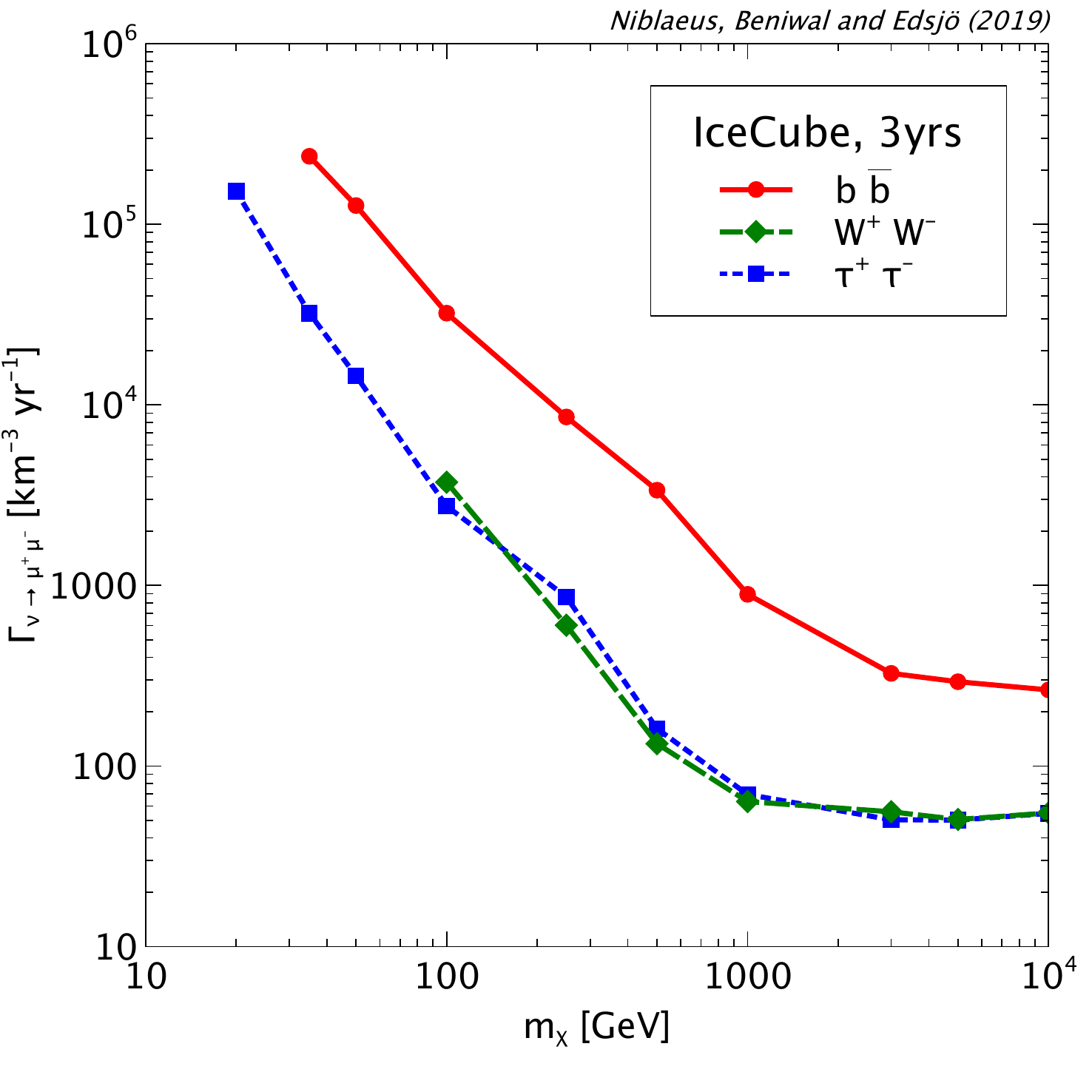}
    \includegraphics[width=0.49\textwidth]{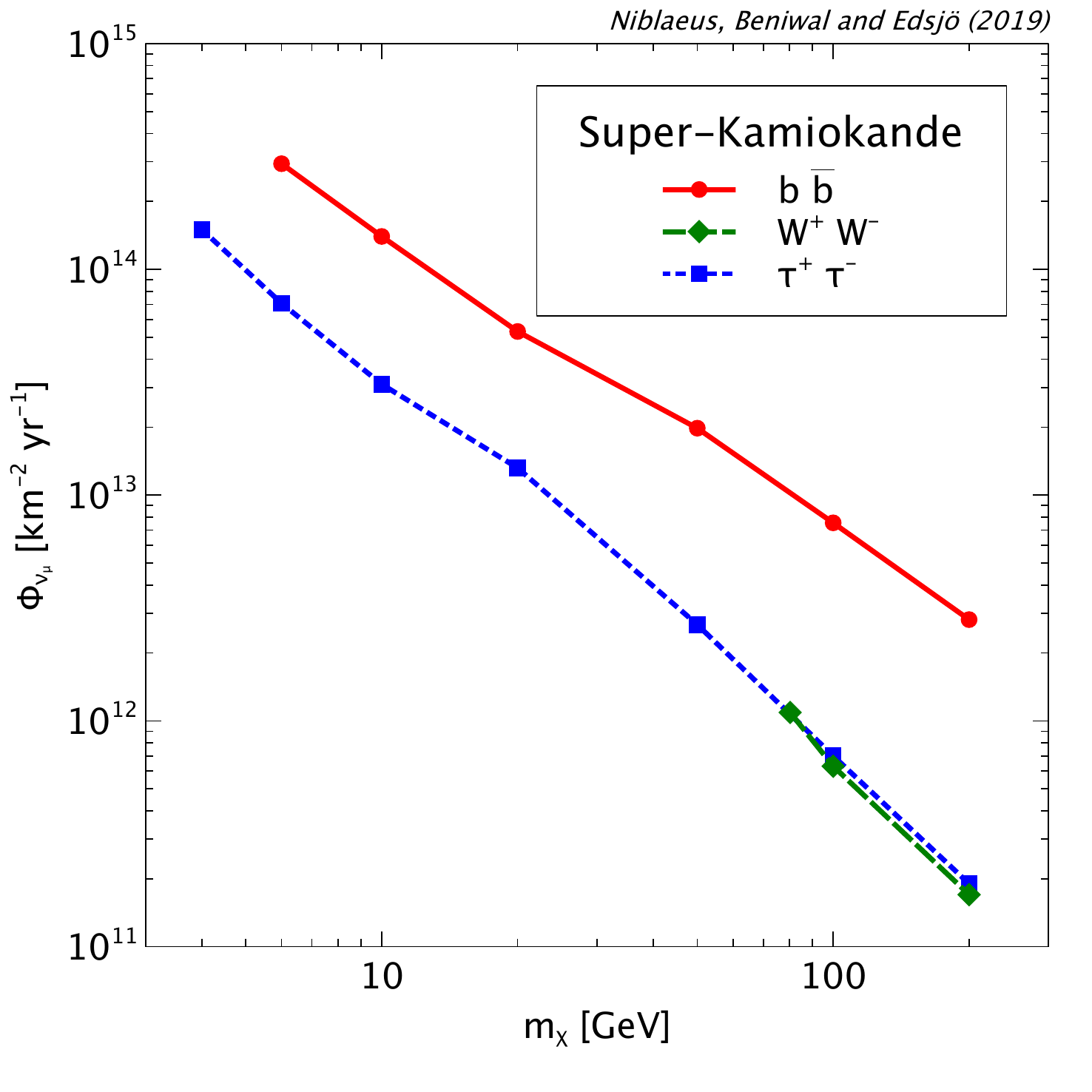}
    \caption{Upper limits on the neutrino-to-muon conversion rate $\Gamma_{\nu \to \mu^+ \mu^-}$ from IceCube \cite{Aartsen:2016zhm} (\emph{left panel}) and on the muon neutrino flux $\Phi_{\nu_\mu}$ from the low-mass analysis of Super-Kamiokande \cite{Choi:2015ara} (\emph{right panel}), both versus the DM mass and for generic annihilation channels.}
    \label{fig:icsk-limits}
\end{figure}

To use the latest results, we will instead develop a method to derive approximate limits on more general DM models using the published results for generic WIMP models.~We first note that our spectra from mediator decays will not be dramatically different than the generic WIMP models in most cases considered in refs.~\cite{Aartsen:2016zhm,Choi:2015ara}.~However, especially for larger $\gamma L$, our spectra will be harder (due to less absorption), thus we cannot use the results presented in the mass-channel plane (e.g.,\ table 4 in ref.~\cite{Aartsen:2016zhm}) directly.~In figure~\ref{fig:icsk-limits}, we reproduce the limits on 
the neutrino-to-muon conversion rate $\Gamma_{\nu \to \mu^+ \mu^-}$ (including both neutrinos and antineutrinos) from IceCube \cite{Aartsen:2016zhm}\footnote{In this process, we found inconsistencies in table 4 of ref.~\cite{Aartsen:2016zhm}. We have received updated values for the limits from the IceCube collaboration. These are shown in figure~\ref{fig:icsk-limits} and are the ones we will use in our analysis.} and on the muon neutrino flux $\Phi_{\nu_\mu}$ (not including anti-neutrinos) from the low-mass analysis of Super-Kamiokande \cite{Choi:2015ara}. Here $\Phi_{\nu_\mu}$ is defined as the flux of muon neutrinos per area, whereas $\Gamma_{\nu \to \mu^+ \mu^-}$ is defined as the number of muons and antimuons produced per volume element from neutrino and anti-neutrino nucleon interactions. In these figures, it is evident that the limits for a given mass are quite dependent on the annihilation channel (i.e.,\ on the actual shape of the neutrino spectrum).~This makes it difficult to use these limits for more general DM models. Here we will try to map these limits onto another quantity that contains information about the (shape of the) spectrum in a better way, i.e.,\ we need a \emph{normalising quantity} that contains information about the spectrum in a more general way than the mass and annihilation channel.

To find this normalising quantity, we first note that what we are really after is a quantity that contains the shape of the neutrino spectrum, or rather how detectable it is in a neutrino telescope. That is, for a given total neutrino flux, how many events do we actually expect?~A harder spectrum will be easier to detect than a softer one, even if the total number of neutrinos would be the same and hence, the limit on the harder neutrino flux would be stronger than on the softer one.~The detectability of a neutrino is roughly proportional to the energy of the neutrino squared, $E_\nu^2$. Here one factor of $E_{\nu}$ comes from the neutrino-nucleon cross section being proportional $E_{\nu}$, and the other comes from the muon range being proportional to the muon energy and hence also to $E_{\nu}$. We then define the following normalising quantity
\begin{equation}\label{eq:emom2}
    \langle E_{\nu_\mu}^2 \rangle \equiv \frac{\int_{E_\textrm{min}}^{m_\chi} E_{\nu_\mu}^2 \left( d\Phi_{\nu_\mu}/dE
    \right) \, dE}{\int_{E_\textrm{min}}^{m_\chi} \left( d\Phi_{\nu_\mu}/dE 
\right) \, dE}
\end{equation}
that we will use for Super-Kamiokande. For IceCube, we can actually do a little bit better as they publish the effective area of the detector as a function of energy \cite{Aartsen:2016zhm}.~Thus, we define the following normalising quantity
\begin{equation}\label{eq:abar}
    \bar{A} \equiv \frac{\int_{E_\textrm{min}}^{m_\chi} A_{\rm eff} (E) \left( d\Phi_{\nu_\mu}/dE + d\Phi_{\ovr{\nu}_{\mu}}/dE \right) \, dE}{\int_{E_\textrm{min}}^{m_\chi} \left( d\Phi_{\nu_\mu}/dE + d\Phi_{\ovr{\nu}_{\mu}}/dE \right) \, dE}.
\end{equation}
In both cases, we will use $E_\textrm{min}=1$ GeV. Note that in eq.~\eqref{eq:emom2}, we only include muon neutrinos, whereas in eq.~\eqref{eq:abar}, we also include muon anti-neutrinos. This is because Super-Kamiokande gives results only for muon neutrinos, whereas IceCube gives results on the sum of muon neutrinos and muon anti-neutrinos.

\begin{figure}[t]
    \centering
    \includegraphics[width=0.49\textwidth]{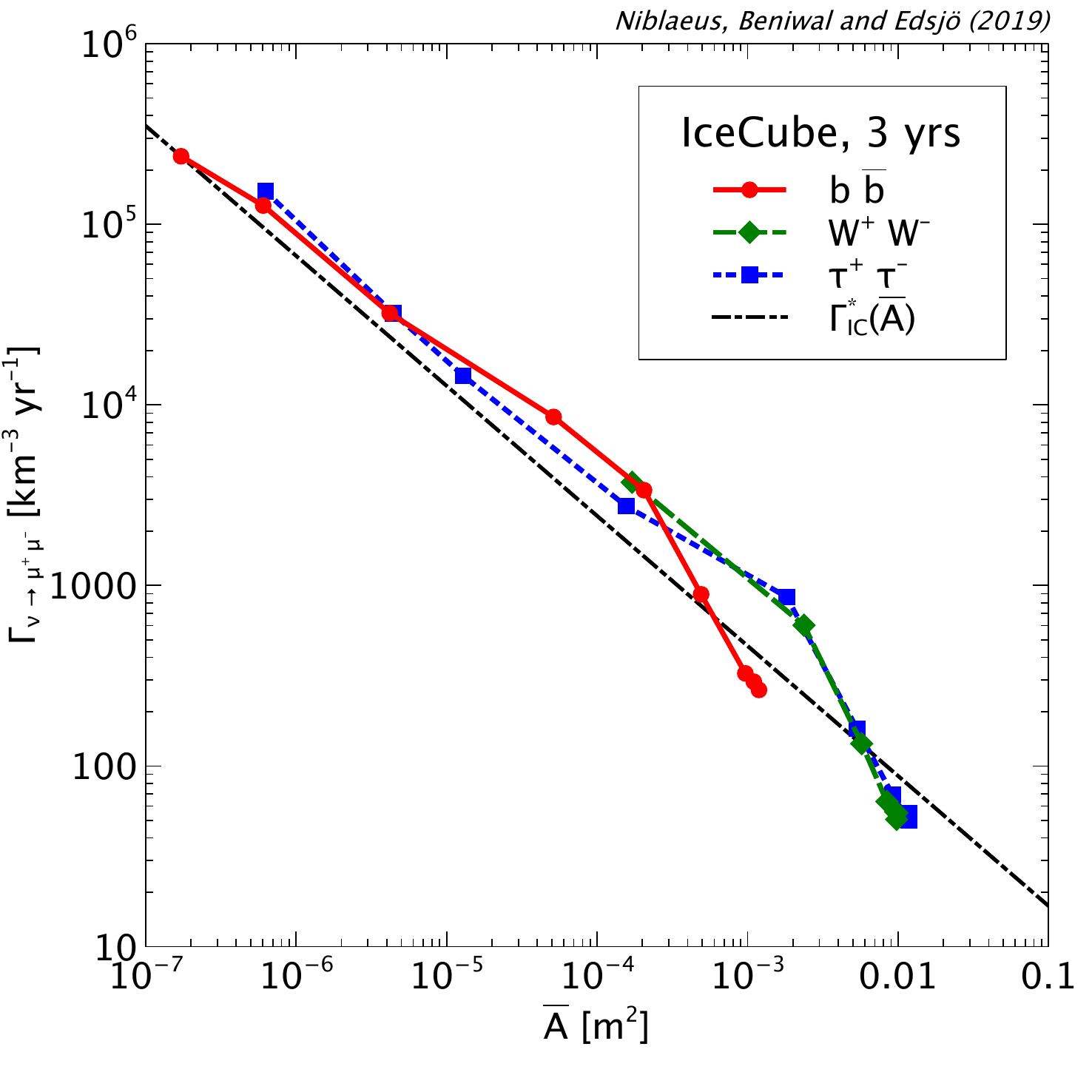}
    \includegraphics[width=0.49\textwidth]{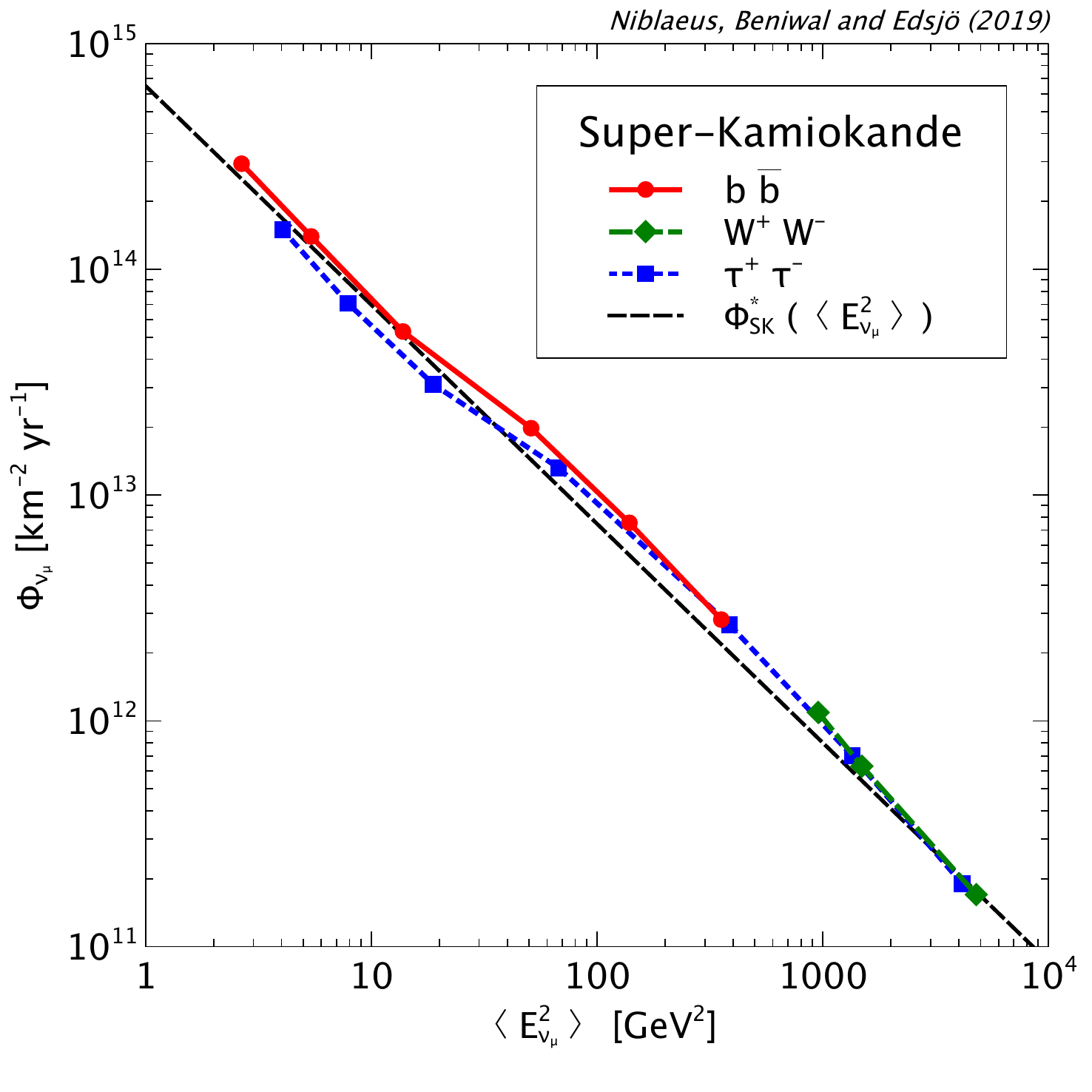}
    \caption{Upper limits on the neutrino-to-muon conversion rate $\Gamma_{\nu \to \mu^+ \mu^-}$ from IceCube \cite{Aartsen:2016zhm} versus $\bar{A}$ (\emph{left panel}) and on the muon neutrino flux from the low-mass analysis of Super-Kamiokande \cite{Choi:2015ara} versus $\langle E_{\nu_\mu}^2\rangle$ (\emph{right panel}). The black dashed lines ($\Gamma^*_{\textrm{IC}} (\bar{A})$ and $\Phi^*_{\textrm{SK}} (\langle E_{\nu_{\mu}}^2 \rangle)$ respectively) correspond to our fit functions for the limit curves (see text for more details).}
    \label{fig:icsk-normlimits}
\end{figure}

In figure~\ref{fig:icsk-normlimits}, we show the same limits as in figure~\ref{fig:icsk-limits}, but now versus our normalising quantities $\bar{A}$ for IceCube and $\langle E_{\nu_\mu}^2\rangle$ for Super-Kamiokande.~We can clearly see that with this procedure, we have managed to reduce the spread of upper limits for different masses and annihilation channels.

We have fitted the limit curves in figure~\ref{fig:icsk-normlimits} with the following functions:
\begin{equation}\label{eq:fit_funcs}
    \Gamma^{*}_{\textrm{IC}}(\bar{A}) = \frac{3.2 \, \textrm{km}^{-3} \,\textrm{yr}^{-1}} {\left(\bar{A}/{\textrm{m}^2}\right)^{0.72}}, \quad
    \Phi^{*}_{\textrm{SK}} (\langle E_{\nu_\mu}^2 \rangle) = \frac{6.5 \times 10^{14} \, \textrm{km}^{-2}\, \textrm{yr}^{-1}}{\left(\langle E_{\nu_\mu}^2 \rangle/{\textrm{GeV}^2}\right)^{0.97}}.
\end{equation} 
For Super-Kamiokande, we expect the fitting function to go as $\langle E_{\nu_\mu}^2\rangle^{-1}$ (as the limits should be proportional to the inverse of the number of detectable events) and this is very close to what we see.~We use the exponent $-0.97$ instead of $-1$ as it gives a slightly better fit. For IceCube, there is still some spread. In this case, we will also need to extrapolate to higher $\bar{A}$-values (as our spectra for large DM masses is harder than the hardest one simulated in ref.~\cite{Aartsen:2016zhm}).~Again, we expect the limits on the neutrino flux to be inversely proportional to the number of detectable events.~Here we instead use the limits on the neutrino-to-muon conversion rate. We have verified by simulations that the neutrino-to-muon conversion rate $\Gamma_{\nu \to \mu^+ \mu^-} \propto (\Phi_{\nu_\mu} + \Phi_{\bar{\nu}_\mu})\bar{A}^{0.28}$, hence we expect the limits on $\Gamma_{\nu \to \mu^+ \mu^-}$ to go as $\bar{A}^{-0.72}$ which is the fitting function we have used in eq.~\eqref{eq:fit_funcs}. We have also verified that if we instead convert the limits on the neutrino-to-muon conversion rate in ref.~\cite{Aartsen:2016zhm} to limits on the neutrino flux and use a fitting function that goes as $\bar{A}^{-1}$, we would get very similar results. 

We use these fitting functions to derive approximate limits on more general DM models in the following way:
\begin{enumerate}
    \item For a more general neutrino spectrum from DM annihilation, calculate $\bar{A}$ and $\langle E_{\nu_\mu}^2\rangle$.
    \item Using the functions in eq.~\eqref{eq:fit_funcs}, calculate the approximate limits on the neutrino-to-muon conversion rate $\Gamma^*_{\textrm{IC}}(\bar{A})$ for IceCube and on the muon neutrino flux $\Phi^*_{\textrm{SK}} (\langle E_{\nu_\mu}^2\rangle)$ for Super-Kamiokande. 
    \item Calculate the total neutrino-to-muon conversion rate $\Gamma_\textrm{tot}$ and the flux of  neutrinos $\Phi_\textrm{tot}$ for our DM model,
    \begin{eqnarray}
        \Gamma_\textrm{tot} & = & \int_{E_\textrm{min}}^{m_\chi} \left(
        \frac{d\Gamma_{\nu \to \mu^+ \mu^-}}{dE} \right)
\, dE, \\[1.5mm]
        \Phi_\textrm{tot} &= & \int_{E_\textrm{min}}^{m_\chi} \left( \frac{d\Phi_{\nu_\mu}}{dE} \right) \, dE,
    \end{eqnarray}
    where $\Gamma_\textrm{tot}$ is measured in km$^{-3}$ ann.$^{-1}$, whereas $\Phi_\textrm{tot}$ is measured in km$^{-2}$ ann.$^{-1}$.
    \item Calculate the upper limits on the DM annihilation rate as
    \begin{eqnarray}
        \Gamma_{A,\;\textrm{IC}}^\nu\;[\textrm{ann.}\,\textrm{s}^{-1}] & = & \frac{1}{3.15 \times 10^7 \;\textrm{s}\,\textrm{yr}^{-1}} \frac{\Gamma^*_{\textrm{IC}} \;[\textrm{km}^{-3}\,\textrm{yr}^{-1}]}{\Gamma_\textrm{tot}\; [\textrm{km}^{-3}\,\textrm{ann.}^{-1}]}, \\[1.5mm]
        \Gamma_{A,\;\textrm{SK}}^{\nu}\;[\textrm{ann.}\,\textrm{s}^{-1}] & = & \frac{1}{3.15 \times 10^7 \;\textrm{s}\,\textrm{yr}^{-1}} \frac{\Phi^*_{\textrm{SK}} \;[\textrm{km}^{-2}\,\textrm{yr}^{-1}]}{\Phi_\textrm{tot}\;[\textrm{km}^{-2}\,\textrm{ann.}^{-1}]}.
    \end{eqnarray}
\end{enumerate}
These correspond to the upper limits shown in figure~\ref{fig:ga-vs-mx}.~In figure~\ref{fig:ga-vs-gl}, we instead show the minimum of these two limits, i.e.,\ the most constraining one.

As given by the data points and our fitted curve in figure~\ref{fig:icsk-normlimits}, we expect our method to be good to about a factor of 2 for IceCube and about 25\% for Super-Kamiokande. This is good enough for the estimates we need to do in this paper.

\noindent \emph{Remark:} For the IceCube analysis, we could have used the results in ref.~\cite{Aartsen:2016exj} which allows for a more accurate analysis of generic DM models.~For the scope of our paper, the more approximate method employed here is good enough, and it also allows us to use the more recent IceCube 3-year data in ref.~\cite{Aartsen:2016zhm}.~We don't expect the results to be significantly different if we instead use the results in ref.~\cite{Aartsen:2016exj}.

\bibliographystyle{JHEP}
\bibliography{med_dec}

\end{document}